\documentclass[]{spie}  

\usepackage{amsmath,amsfonts,amssymb}
\usepackage{graphicx}
\usepackage[colorlinks=true, allcolors=blue]{hyperref}
\usepackage{longtable}
\usepackage{booktabs}   

\title{The Evolution of the ACIS Contamination Layer on the Chandra X-ray Observatory from 2010 to 2026}

\author[a]{Paul P. Plucinsky}
\author[a]{Peter W. Ratzlaff}
\author[a]{Akos Bogdan}
\author[b]{Herman L. Marshall}
\affil[a]{Smithsonian Astrophysical Observatory, Center for Astrophysics $\vert$ Harvard \& Smithsonian, MS-3, 60 Garden
St., Cambridge, MA 02138, USA}
\affil[b]{MIT Kavli Institute for Astrophysics and Space Research,
Cambridge, MA 02139, USA}

\authorinfo{Further author information: (Send correspondence to
  P.P.P)\\P.P.P.: E-mail: pplucinsky@cfa.harvard.edu}

\newcommand{\etal}{{\em et al.}}

\DeclareRobustCommand{\ion}[2]{\textup{#1\,\textsc{\lowercase{#2}}}}
 
\begin{document} 
\maketitle

\begin{abstract}

  The {\it Chandra X-ray Observatory} (CXO) was launched over 27 years ago and has been 
delivering spectacular science over the course of its mission.  The 
{\it Advanced CCD Imaging Spectrometer}
(ACIS) is the prime instrument on the satellite, 
conducting over 90\% of the observations. The CCDs operate at a temperature
of $-$120$\rm{^\circ C}$ and the optical blocking filter (OBF) in front of the CCDs is at a
temperature of approximately $-60 \rm{^\circ C}$.  The surface of the OBF has 
accumulated a layer of contamination over the course of the mission, as it is the coldest
surface exposed to the interior to the spacecraft.  We have been 
characterizing the thickness, chemical composition, and spatial distribution of the contamination
layer as a function of time over the mission.
The contamination model has required several revisions over the course of the mission as the properties of the contamination layer have changed and our understanding of the layer has improved.
In this paper, we evaluate the performance of the current contamination model (N0016 released in CalDB~4.12.3 on 16 December 2025) using the most recent calibration observations conducted from 2023 to 2026 by using the standard model spectrum for the supernova remnant 1E~0102.2-7219 (E0102) developed by the {\em International Astronomical Consortium for High Energy} (IACHEC), spectral data from the cluster of galaxies known as Abell~1795, and high resolution X-ray spectra of Mrk 421.
This evaluation has been complicated by the decreasing observed counts at low energies, especially the \ion{O}{vii}~He$\alpha$~line complex and the \ion{O}{viii}~Ly$\alpha$ line from E0102 which are no longer useful for this purpose.   The analyses of the E0102, Abell~1795, and Mrk~421 data show that the
current model of the contamination adequately predicts the additional absorption through mid-2026.

\end{abstract}


\section{INTRODUCTION}
\label{sec:intro}  


The {\it Chandra X-ray Observatory} (CXO) was launched on 23 July 1999
on the Space Shuttle {\it Columbia}.  An overview of the mission and
its instruments are presented in 
Weisskopf~\etal~(2000)~\cite{weisskopf2000} and an update on
the mission was provided in Weisskopf~\etal~(2012)\cite{weisskopf2012}.  
The CXO carries two imaging instruments, the {\it Advanced CCD Imaging
Spectrometer} (ACIS) discussed in Garmire~\etal~(1992)~\cite{garmire92}
and Garmire~\etal~(2003)~\cite{garmire03} and the {\it High
Resolution Camera} (HRC) discussed in
Murray~\etal~(1997)~\cite{murray1997}.  
In addition, the CXO
carries two gratings instruments known as the {\it High Energy
  Transmission Grating} (HETG) described in 
Canizares~\etal~(2000)~\cite{canizares2000} and the {\it Low
  Energy Transmissions Grating} (LETG) 
Brinkman~\etal~(2000)~\cite{brinkman2000}.

ACIS is the primary detector system on the CXO with a nominal bandpass of
 0.3--10.0~keV, conducting over 90\% of the
observations.  ACIS contains 10 CCDs arranged into two arrays. One
array, the ACIS Imaging array (ACIS-I), consists of four 
front-illuminated (FI) CCDs arranged in a $2\times2$ array, and the other
array, the ACIS Spectroscopy array (ACIS-S), consists of  four FI CCDs
and two back-illuminated (BI) CCDs arranged
in a $1\times6$ array. Both the ACIS-I and ACIS-S arrays are used for imaging spectroscopy. In addition, ACIS-S is also used as the the readout
detector for the HETG and LETG.  
The BI CCDs have higher quantum efficiency at low energies than the FI CCDs and are therefore preferred over the FI CCDs for some imaging observations.  The BI CCD in the ACIS-S array that is positioned at the aimpoint is labelled ``S3'' and the FI CCD in the ACIS-I array that is positioned at the aimpoint is labelled ``I3''.

 In order to suppress optical and infrared photons but transmit the
 X-ray photons of interest, both ACIS arrays have {\em Optical
   Blocking Filters} (OBFs) inserted in the optical path.  The filters 
were produced by ${\tt
   Luxel^{TM}}$ and are made of polyimide with Al deposited on both
 sides of the polyimide.  The two filters are of slightly different
 thicknesses: the ACIS-S OBF (OBF-S) is 100/200/30 nm of
 Al/polyimide/Al and the ACIS-I OBF is 130/200/30 nm of
 Al/polyimide/Al. The OBFs are placed about 12~mm in front of the CCDs facing
 the CXO mirrors.  The volume around the CCDs is effectively
 isolated from the interior of the spacecraft, while the surface of
 the OBFs facing the mirrors is exposed to the interior of the
 spacecraft. The CCD focal plane was regulated at a temperature of
 -120~$\rm{^\circ C}$ for most of the mission.  However, as the mission has progressed the focal plane temperature is frequently unregulated, reaching temperatures as high as -105~$\rm{^\circ C}$ for science observations and as high as -90~$\rm{^\circ C}$ during perigee passages.
  The OBFs are positioned at the top of the ACIS Camera Body (CB)
 which was regulated at -60~$\rm{^\circ C}$ early in the mission, but has been
 unregulated from April 2008 fluctuating between -72~$\rm{^\circ C}$ and
 -60~$\rm{^\circ C}$ (see
 Plucinsky~\etal~2016~\cite{plucinsky2016}  for details). 
 For a pristine filter with no contamination layer, the centers of the
 filters are warmer by $\sim2-4$ $\rm{^\circ C}$ due to the radiative heat load
 of the warm mirrors (+20~$\rm{^\circ C}$) and the optical bench assembly. 

It was noticed early in the mission\cite{plucinsky2003} that the low energy sensitivity of the ACIS instrument was decreasing with time. It was quickly determined that this loss of detection efficiency was the result of a contamination layer building up on the surface of the OBFs facing the spacecraft interior.  The contamination layer continues to accumulate even after 27 years on orbit.  The accumulation rate, the chemical composition, and the spatial distribution of the contaminant have all varied with time over the mission. The accumulation rate exhibited a steep rise at the beginning of the mission, a flattening from 2003 until 2010, and then another steep rise from 2010 onward. We reported in 2016\cite{plucinsky2016} on our efforts to reduce the accumulation rate by turning on the ACIS Detector Housing (DH) heater which regulates the CB and hence OBF edges at -60~$\rm{^\circ C}$.  There was no measurable effect on the accumulation rate due to the DH heater regulating the CB at -60~$\rm{^\circ C}$. Therefore, we decided to turn the DH heater off again to provide more margin on the FP temperature. A series of papers\cite{odell2013,odell2015,odell2017} were written that present the results of simulations of a possible bakeout of the ACIS instrument to remove the contamination layer.  The reader is referred to those papers for a detailed description of the physical and thermal properties of the ACIS instrument relevant for a bakeout.

 The {\it Chandra X-ray Center} (CXC) calibration team regularly updates the contamination model when the calibration observations indicate that the model no longer adequately describes the absorption produced by the contamination layer.  We have written a series of papers that documents the major changes to the contamination model with time.
 In our 2018 paper\cite{plucinsky2018}, we reported that the N0010 contamination model was significantly over-estimating the thickness of the layer on the ACIS-I array OBF at that point in time.  The N0011 contamination model was released soon after that analysis in June 2018 which altered the time dependence just for the ACIS-I OBF.  A later release, N0012, in October 2018 modified the time-dependence for the ACIS-S OBF and also included a more complicated absorption model for the O and F edges. 
A further revision was made in November 2019 with the N0013 model to increase the accumulation rate once it was identified that the N0012 model was under-estimating the absorption in the calibration observations acquired in 2019.
The N0012 model contained a temporal model with the accumulation rate decreasing after 2018, while the 2019 calibration data showed that the accumulation had been increasing roughly linearly over that period.
In our 2020 paper\cite{plucinsky2020}, we showed that the N0013 model was under-predicting the contamination layer based on the 2020 calibration data.  The CXC released the N0014 contamination model in December 2020 which further adjusted the accumulation rate to be closer to a linear extrapolation.  In our 2022 paper\cite{plucinsky2022}, we showed that the N0014 contamination model was under-predicting the optical depth by 17~\% at 0.66~keV and by 10~\% at 1.0~keV.  The N0015 contamination model increased the accumulation rate at late times and was released in CalDB~4.10.2 on 15 November~2022.
In this paper, we compare the 2023-2026 calibration data with the N0016 model, released in 2025.

\section{UPDATES IN THE N0016 MODEL}
\label{sec:n0016}  

The structure of the contamination model has been discussed in detail in our previous papers\cite{marshall2004,plucinsky2018}.  In brief, the contamination model now has five components each with its own time dependence at both the center and edge of the detector. One component is intended to model the C-K absorption edge.  The other four components are intended to model the O and F absorption, each with a Henke edge structure as one component and Henke edge with near edge structure as another component.  K-shell and L-shell absorption are both included in the absorption models.  The spatial shape is the same for all components; only the time dependences of the spatial components' normalizations differ.
Compared to the models released as versions N0014 and N0015, the latest release changes the time dependences of the optical depths of the contaminant components, as shown in sections \ref{sec:a1795} and \ref{sec:Mrk421}.  The model of the absorption edges and the variation with detector position (the spatial model) were not changed.

\section{EVALUATION OF THE N0016 MODEL WITH E0102}
\label{sec:e0102}  

The {\it Chandra X-ray Center} (CXC) calibration team conducts regular observations throughout the year to characterize the changes in the contamination layer.  A bright continuum source is used with the LETG and ACIS to provide sufficient counts to measure small changes in the C, O, and F edges.  Since 2014, the blazar Mrk~421 has been used given its relatively high flux in the so-called ``Big Dither'' mode which sweeps the source spectrum over a large fraction of the ACIS-S array in a single observation.  The cluster of galaxies known as Abell~1795 (A1795) is used as a bright and constant extended source to map the contamination layer on large regions on the ACIS-S and ACIS-I array.  Finally, the Small Magellanic Cloud supernova remnant 1E~0102.2-7219 (E0102) is used to provide a constant, line-dominated source in the low energy part of the bandpass (0.5-1.5~keV). We have developed a spectral model for E0102 as part of the efforts of the {\em International Astronomical Consortium for High Energy Calibration} (IACHEC) to develop models of standard candles\cite{plucinsky2017} for X-ray astronomy. The IACHEC model is a high-fidelity model that attempts to accurately represent the flux from E0102 in the 0.3-1.6~keV bandpass, but it should not be regarded as the absolute truth.  The IACHEC model does facilitate a comparison of line fluxes as a function of time since E0102 does not vary significantly in time.  If the contamination model is correctly accounting for the growth of the contamination layer, the E0102 line fluxes and the A1795 fluxes in narrow energy bands should be constant in time.  We used the standard tools in the {\em {Chandra Interactive Analysis of Observation}} (CIAO)\cite{fruscione2026} software to generate response matrix files (rmfs) and ancillary response files (arfs).  We used CIAO version 4.18.0 and CXC Calibration Database (CALDB) 4.12.3 for this analysis.  The contamination file in the CXC Calibration Database (CALDB) at the time this paper was written is called {\tt acisD1999-08-13contamN0016.fits} and we selected the option to generate rmfs for a focal plane temperature of -120~C.

\begin{longtable}{rcrrcrrcrc} 
\caption{ACIS S3 Observations of E0102} 
\label{tab:s3obs} \\

\toprule
ObsID	& Date	   & ChipX  & ChipY	& Node	& Exposure	& Counts	& Frame	& 1stRow & Nrows \\
	&	   & 	    &           &	& (s)		& (0.3-2keV)    & (s)   &         &  \\
\midrule
\endfirsthead

\caption[]{ACIS S3 Observations of E0102 -- continued} \\

\toprule
ObsID	& Date	   & ChipX  & ChipY	& Node	& Exposure	& Counts	& Frame	& 1stRow & Nrows \\
	&	   & 	    &           &	& (s)		& (0.3-2keV)    & (s)   &         &  \\
\midrule
\endhead

\bottomrule
\endfoot

 3545	& 2003.60  & 357.6  & 496.8	& 1	& 7863.8	&   57133	&1.1	&385	&256	\\
 6765	& 2006.22  & 106.8  & 498.8	& 0	& 7636.7	&   51768	&0.8	&384	&256	\\
 8365	& 2007.11  & 100.2  & 494.3	& 0	&20985.3	&  138698	&0.8	&384	&256	\\
 9694	& 2008.10  & 100.1  & 492.8	& 0	&19196.5	&  124804	&0.8	&384	&256	\\
10654	& 2009.17  &  98.9  & 495.2	& 0	& 7307.1	&   45532	&0.8	&335	&256	\\
10655	& 2009.17  &  98.2  & 493.0	& 0	& 6810.7	&   43234	&0.4	&433	&128	\\
10656	& 2009.18  & 342.0  & 494.3	& 1	& 7763.9	&   48603	&0.8	&335	&256	\\
11957	& 2009.99  & 103.2  & 491.1	& 0	&18447.8	&  112437	&0.8	&335	&256	\\
13093	& 2011.08  &  89.8  & 487.9	& 0	&19049.4	&  108275	&0.8	&335	&256	\\
14258	& 2012.03  &  86.1  & 493.7	& 0	&19049.3	&  102049	&0.8	&360	&256	\\
15555	& 2013.03  & 676.5  & 489.5	& 2	&23837.3	&  114679	&0.8	&360	&256	\\
15558	& 2013.06  & 854.4  & 479.7	& 3	&23051.7	&  107679	&0.8	&360	&256	\\
15556	& 2013.07  & 666.1  & 164.2	& 2	&23841.3	&   94733	&0.8	& 42	&256	\\
15467	& 2013.08  &  91.7  & 488.9	& 0	&19082.2	&   92610	&0.8	&360	&256	\\
15557	& 2013.09  & 657.9  & 917.2	& 2	&24191.7	&   78972	&0.8	&769	&256	\\
15559	& 2013.09  & 854.8  & 162.4	& 3	&23842.1	&   92216	&0.8	& 42	&256	\\
16589	& 2014.24  &  78.1  & 485.5	& 0	& 9569.5	&   40194	&0.8	&360	&256	\\
17380	& 2015.16  & 119.2  & 489.0	& 0	&17655.1	&   65808	&0.8	&360	&256	\\
17381	& 2015.18  & 659.7  & 165.6	& 2	& 9573.6	&   27095	&0.8	& 42	&256	\\
17382	& 2015.18  & 671.5  & 926.7	& 2	& 9572.7	&   21161	&0.8	&769	&256	\\
17688	& 2015.54  & 110.7  & 481.6	& 0	& 9569.6	&   33973	&0.8	&360	&256	\\
17689	& 2015.54  & 661.3  & 158.0	& 2	& 9573.5	&   25482	&0.8	& 42	&256	\\
17690	& 2015.54  & 660.8  & 914.2	& 2	& 9572.7	&   20772	&0.8	&769	&256	\\
18418	& 2016.21  &  99.0  & 480.2	& 0	&14326.2	&   45687	&0.8	&360	&256	\\
18420	& 2016.20  & 651.9  & 919.2	& 2	&19085.4	&   36556	&0.8	&769	&256	\\
18419   & 2016.22  & 642.3  & 157.9     & 2     &19084.7        &  45491 	&0.8    & 42    &256     \\
19850   & 2017.22  &  98.6  & 484.9     & 0     &14326.21       &  38782        &0.8    &360    &256     \\
19851   & 2017.23  & 647.6  & 166.6     & 2     &19085.35       &  38603        &0.8    & 42    &256     \\
19852   & 2017.24  & 652.3  & 922.4     & 2     &19085.38       &  30461        &0.8    &769    &256     \\
20639   & 2018.20  & 103.8  & 489.4     & 0     &14326.21       &  32681        &0.8    &360    &256     \\
20640   & 2018.12  & 645.2  & 165.3     & 2     &19013.47       &  32687        &0.8    & 42    &256     \\
20641   & 2018.10  & 658.2  & 923.1     & 2     &19084.61       &  26226        &0.8    &769    &256     \\
21804   & 2019.21  & 95.5   & 484.1  	& 0     &14325.42       &  27154        &0.8    &360    &256    \\
21805   & 2019.19  & 643.0  & 164.8  	& 2     &19085.41       &  27341        &0.8    & 42    &256    \\
21806   & 2019.21  & 649.9  & 918.6  	& 2     &19084.55       &  21416        &0.8    &769    &256    \\
22805   & 2020.10  &  96.5  & 483.9  	& 0     &14326.27       &  23217        &0.8    &360    &256    \\
22806   & 2020.12  & 641.5  & 162.6  	& 2     &18011.87       &  21495        &0.8    & 42    &256    \\
22807   & 2020.12  & 650.1  & 920.5  	& 2     &19084.63       &  17931        &0.8    &769    &256    \\
24577   & 2021.11  & 101.9  & 484.3     & 0     &13851.01       &  18631        &0.8    &360    &256     \\
24578   & 2021.13  & 642.6  & 164.3     & 2     &19085.38       &  18822        &0.8    & 42    &256     \\
24579   & 2021.16  & 649.9  & 922.5     & 2     &18133.35       &  14060        &0.8    &769    &256     \\
25618   & 2022.23  & 89.8   & 486.4     & 0     &14444.94       &  15810        &0.9    &385    &256     \\
25619   & 2022.20  & 634.8  & 163.6     & 2     &17943.83       &  14690        &0.8    & 42    &256     \\
25620   & 2022.21  & 648.1  & 923.3     & 2     &19085.42       &  12081        &0.8    &769    &256     \\
25618   & 2022.23  &  89.8  & 486.4     & 0     &14444.94       &  15813        &0.9    &385    &256     \\
26987   & 2023.22  & 101.0  & 489.2     & 0     &14446.69       &  12938        &0.9    &360    &256     \\
26988   & 2023.22  & 645.5  & 168.7     & 2     & 9559.93       &  6585         &0.8    & 42    &256     \\
26989   & 2023.22  & 653.3  & 926.2     & 2     & 9664.13       &  5131         &0.9    &769    &256     \\
27761   & 2023.22  & 654.2  & 926.3     & 2     & 9664.18       &  5090         &0.9    &769    &256     \\
27760   & 2023.23  & 647.8  & 169.3     & 2     & 9559.93       &  6607         &0.8    & 42    &256     \\
28432   & 2024.37  &  95.2  & 485.8     & 0     &14443.97       &  10084        &0.9    &360    &256     \\
28434   & 2024.39  & 647.1  & 920.4     & 2     &19227.38       &  8143         &0.9    &769    &256     \\
28433   & 2024.40  & 640.0  & 163.9     & 2     &19071.12       &  10265        &0.8    & 42    &256     \\
29586   & 2025.39  & 633.7  & 160.3     & 2     &18120.67       &  7918         &0.8    & 42    &256     \\
29585   & 2025.40  &  86.8  & 482.3     & 0     &14446.77       &  8022         &0.9    &360    &256     \\
29587   & 2025.40  & 637.7  & 917.8     & 2     &18203.20       &  6355         &0.9    &769    &256     \\
31373   & 2026.39  &  80.9  & 484.0     & 0     &14445.86       &  6643         &0.9    &360    &256     \\
31375   & 2026.40  & 634.8  & 918.6     & 2     &19016.87       &  5331         &0.9    &769    &256     \\
31374   & 2026.40  & 624.7  & 160.1     & 2     &20974.25       &  7654         &0.8    & 42    &256     \\
\end{longtable}

\begin{figure} [ht]
   \begin{center}
  \includegraphics[trim={10 45 60 30}, clip, height=6.3cm]{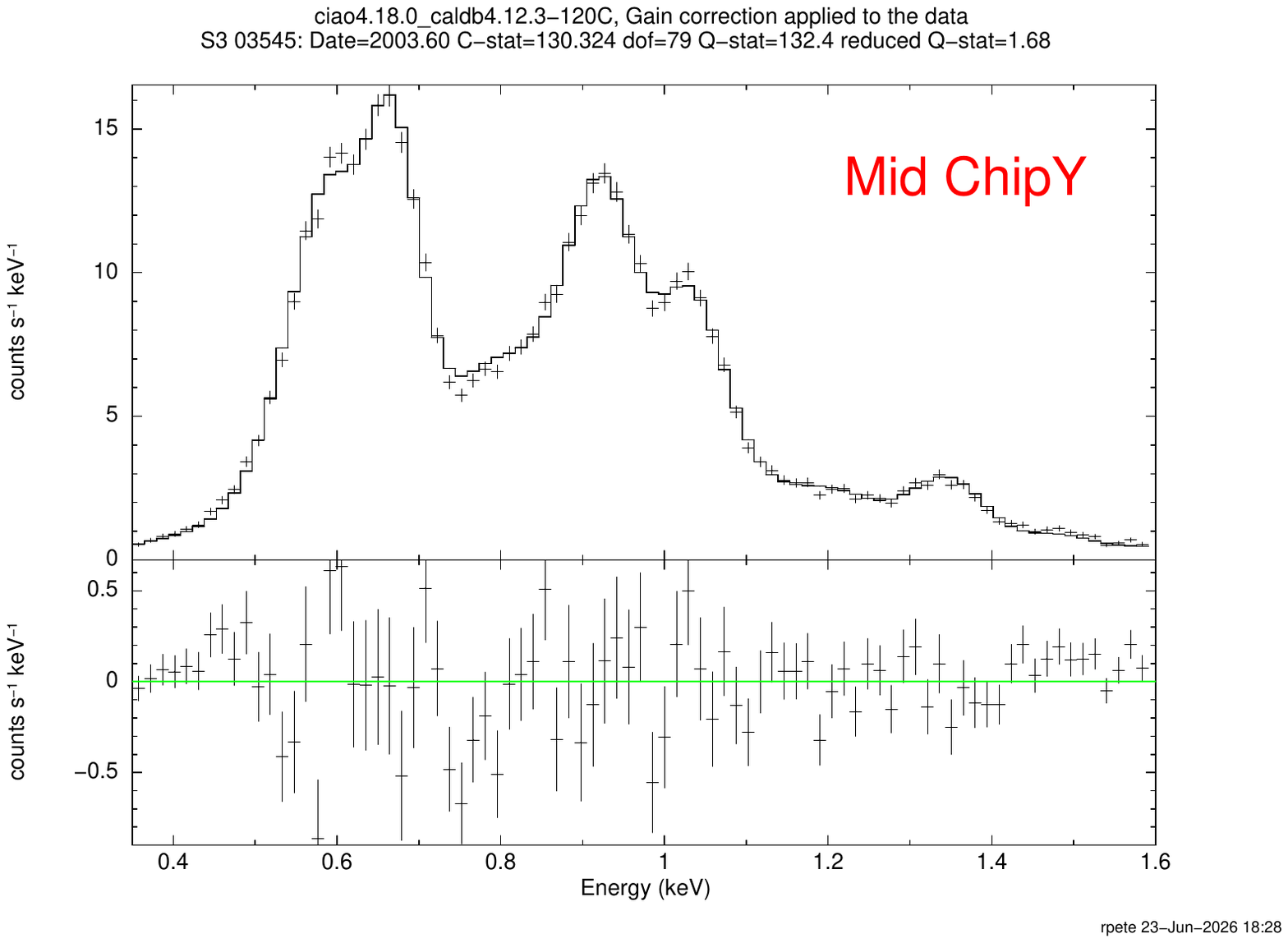}
  \includegraphics[trim={10 45 60 30}, clip, height=6.3cm]{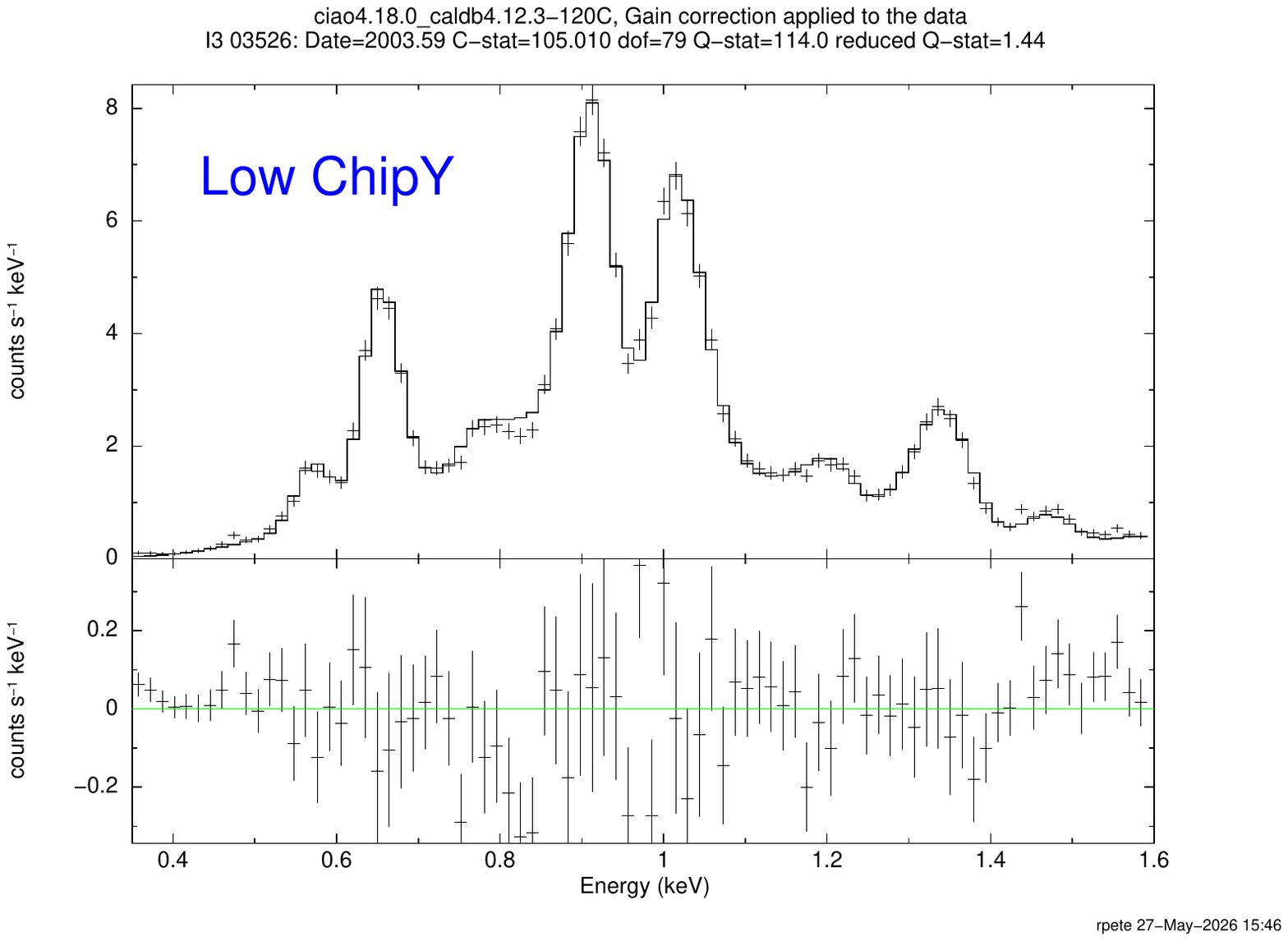}
   \end{center}
   \vspace{-0.2cm}
   \caption[example] 
   { \label{fig:e0102_spectra_1} 
Sample E0102 spectra fit with the IACHEC standard model. LEFT: OBSID 3545, an S3 observation acquired at mid chipy. RIGHT: OBSID 3526, an I3 spectrum acquired at low chipy where the spectral resolution of the FI CCDs is the highest.}
   \end{figure}

\begin{figure} [ht]
   \begin{center}
  \includegraphics[trim={10 45 60 30}, clip, height=6.3cm]{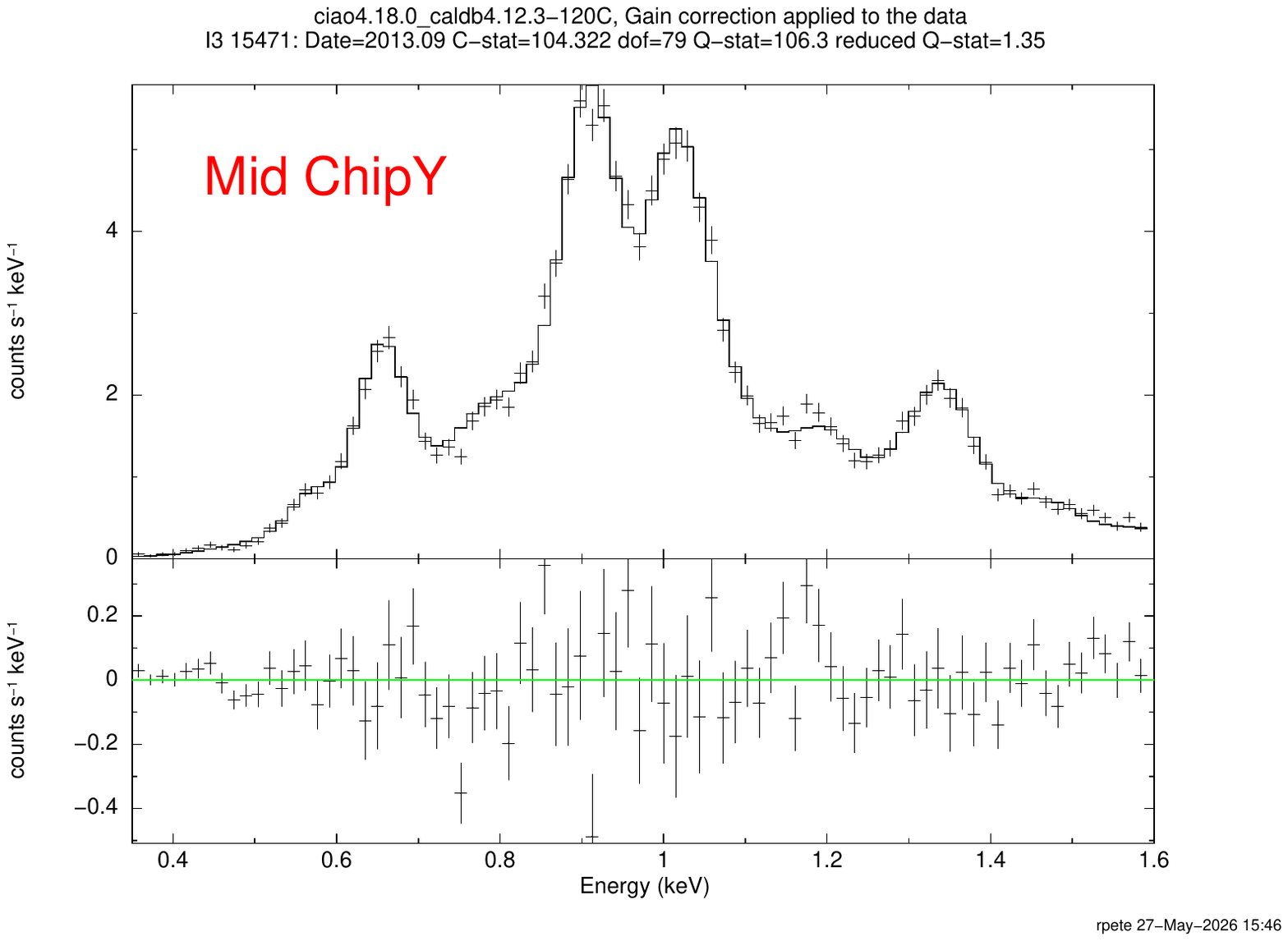}
  \includegraphics[trim={10 45 60 30}, clip, height=6.3cm]{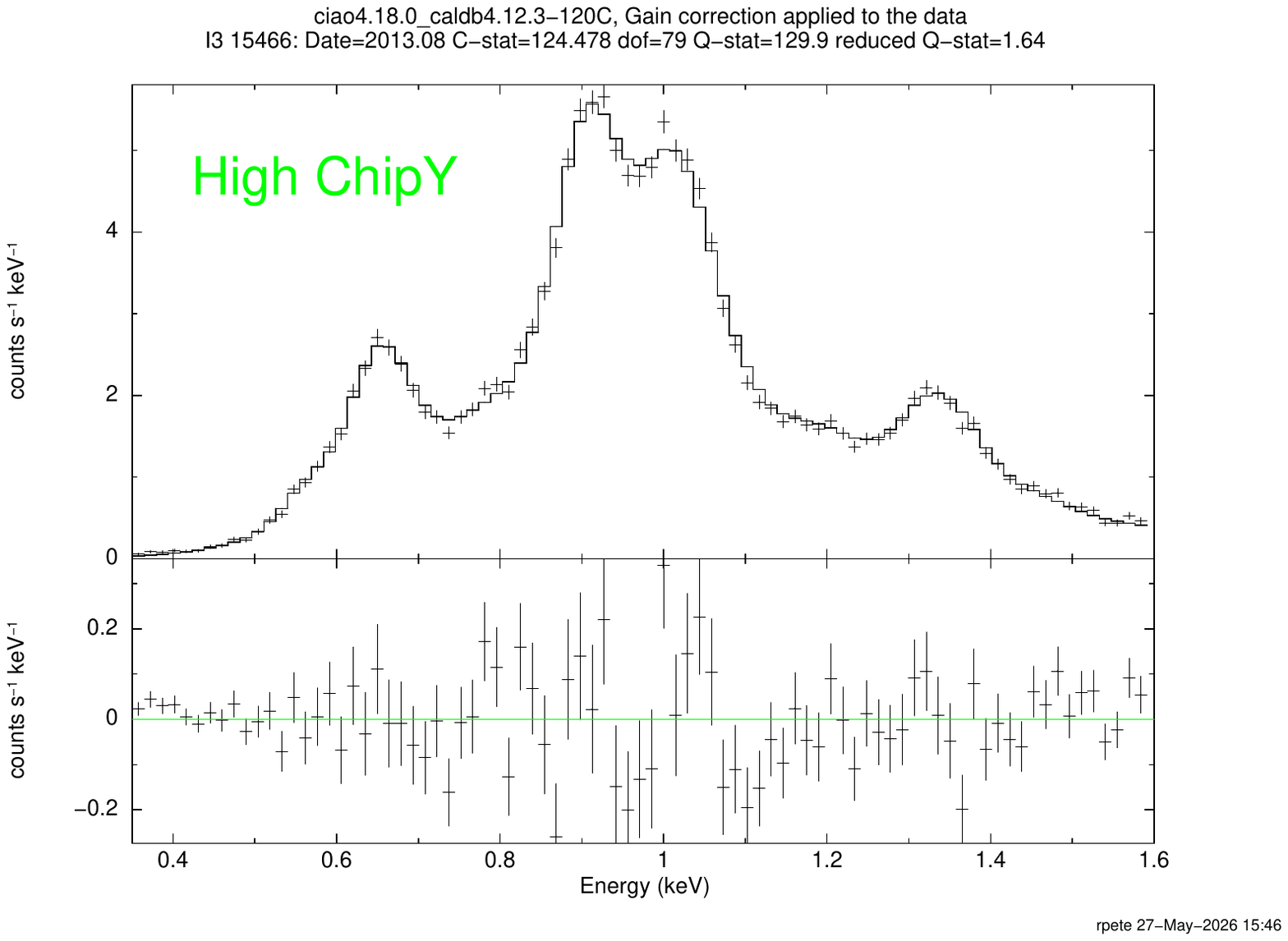}
   \end{center}
   \vspace{-0.2cm}
   \caption[example] 
   { \label{fig:e0102_spectra_2} 
Sample E0102 spectra fit with the IACHEC standard model. LEFT: OBSID 15471, an I3 observation acquired at mid chipy. RIGHT: OBSID 15466, an I3 spectrum acquired at high chipy where the spectral resolution of the FI CCDs is the lowest.}
   \end{figure}

The E0102 observations on S3 and I3 used in this analysis are listed in Tables~\ref{tab:s3obs} and \ref{tab:i3obs} respectively.  We make use of subarray observations of E0102 to reduce pileup in the early mission data when the count rate was higher due to the contamination layer being thinner.  We fit all of these data sets with the IACHEC model for E0102. Although the standard IACHEC model has many parameters, most of them are held fixed when we fit the data for calibration purposes.  The continuum components are fixed and the interstellar absorption components are held fixed.  All the line energies and widths are also held fixed.  Typically, we freeze all line normalizations except for the four normalizations of the brightest lines/line complexes.  We allow the normalizations for the \ion{O}{vii}~He$\alpha$~{\em r} line, the \ion{O}{viii}~Ly$\alpha$ line, the \ion{Ne}{ix}~He$\alpha$~{\em r} line, and \ion{Ne}{X}~Ly$\alpha$ line to vary in the fit.  For the \ion{O}{vii}~He$\alpha$ and  \ion{Ne}{ix}~He$\alpha$ triplets, we link the normalizations of the {\em f}, {\em i}, and {\em r} lines to each other and only allow one of them to vary during the fitting process.  In this way, the triplet can increase or decrease its normalization as a group but the normalizations of the individual lines in the triplet can not vary independently of each other.  There is also a global constant that multiplies the entire spectrum that is allowed to vary.  In this manner, we allow only 5 of the 208 parameters in the IACHEC model to vary when we fit. We are essentially allowing the normalizations of the four brightest line/line complexes to vary while freezing the weaker lines and the continuum to the values in the IACHEC model.  Figure~\ref{fig:e0102_spectra_1} shows spectra from the S3 CCD at mid chipy and the I3 CCD at low chipy from 2003.  The higher spectral resolution of the I3 CCD at low chipy is evident in this figure.  The \ion{O}{vii}~He$\alpha$ triplet is centered at $\sim0.57$~keV, the \ion{O}{viii}~Ly$\alpha$ line is centered at $\sim0.65$~keV, the \ion{Ne}{ix}~He$\alpha$ triplet is centered at $\sim0.92$~keV, and \ion{Ne}{X}~Ly$\alpha$ line is centered at $\sim1.02$~keV.  Note also the \ion{Mg}{xi}~He$\alpha$ triplet at 1.34~keV.
Figure~\ref{fig:e0102_spectra_2} shows spectra from the I3 CCD at mid chipy and at high chipy from 2013.  The difference in spectral resolution on the I3 CCD at the three positions, low, mid, \& high chipy, is evident in these spectra.  The \ion{O}{vii} \& \ion{O}{viii} lines and the 
\ion{Ne}{ix} \& \ion{Ne}{X} lines blend together at high chipy making it more challenging to determine the line normalizations at that position. 

We followed this approach in all of our previous analyses through our 2022 paper.  In 2022, it had become clear that the \ion{O}{vii}~He$\alpha$~{\em r} and the \ion{O}{viii}~Ly$\alpha$ results were becoming less reliable and were not useful for constraining the contamination model due to the low observed flux that results in large statistical errors. Therefore, in this paper we do not report results for the  \ion{O}{vii}~He$\alpha$~{\em r} line after 2021 and the \ion{O}{viii}~Ly$\alpha$ line after 2023. We now include results for the \ion{Mg}{XI}~He$\alpha$~{\em r} line from the beginning of the mission through 2026.  As the contamination layer has continued to accumulate and the absorption has increased, the \ion{Mg}{XI}~He$\alpha$~{\em r} line has become more useful in constraining the contamination model.  The inclusion of the \ion{Mg}{XI}~He$\alpha$~{\em r} line at 1.352~keV has the added benefit that the results can be compared to the Al~K line from the external calibration source (ECS) at the nearby energy of 1.486~keV.  With this modified approach, there are six free parameters in our fits up until 2021, five free parameters from 2021 to 2023, and only four free parameters from 2023 to 2026.  In addition, we included the \ion{Mg}{XI}~He$\alpha$~{\em r} line in the gain adjustment applied to the data and require the gain adjustment to go to zero at 1.6~keV where previously we had required it to go to zero at 1.1~keV.

  \begin{figure} [ht]
   \begin{center}
  \includegraphics[trim={20 150 10 180}, clip, height=6.7cm]{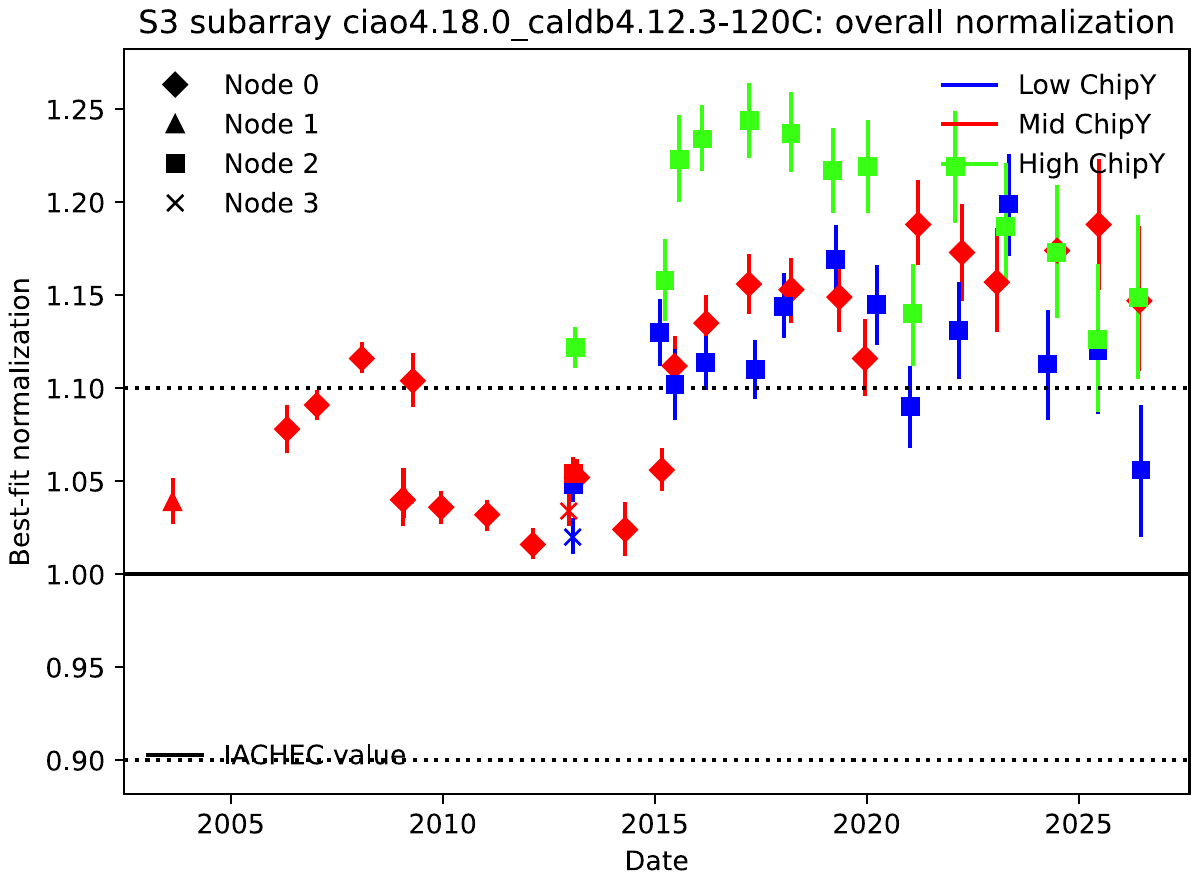}
  \includegraphics[trim={20 150 10 180}, clip, height=6.7cm]{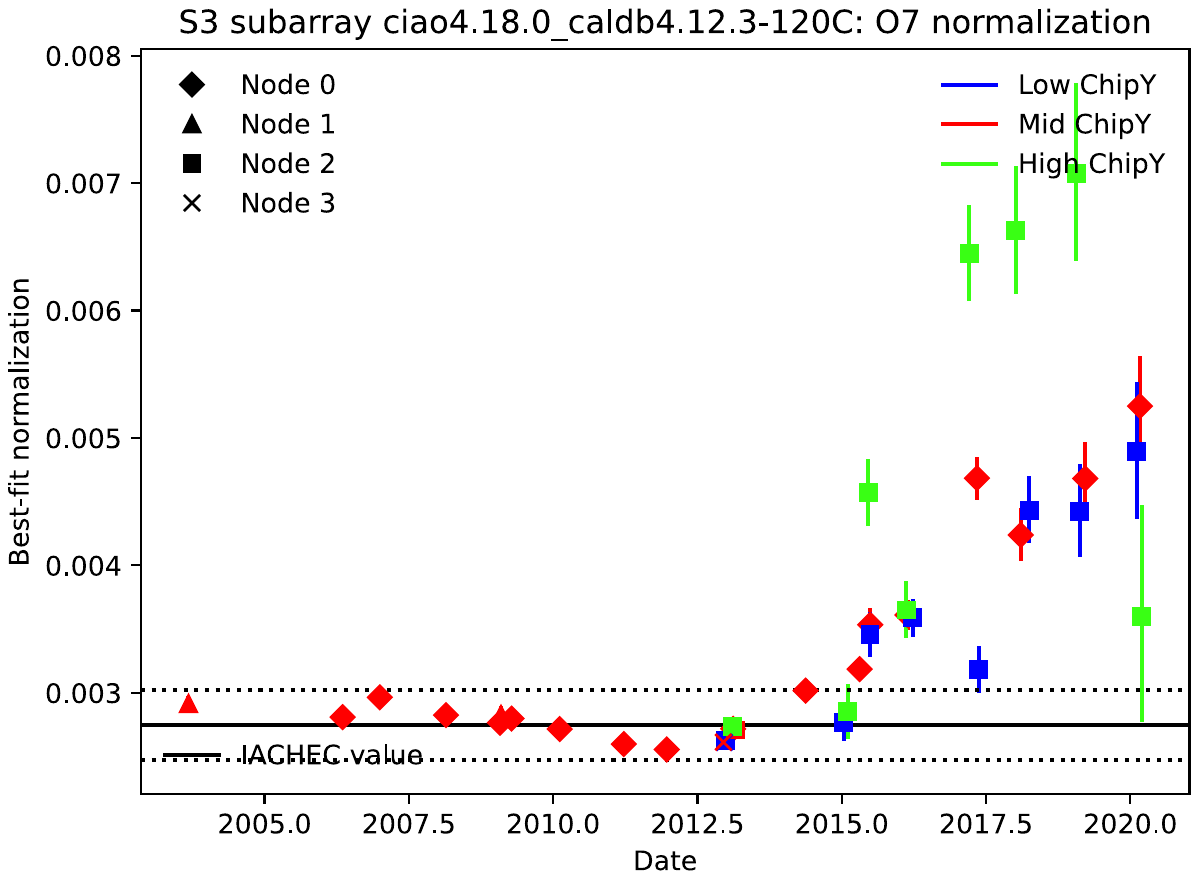}
   \end{center}
   \vspace{-0.8cm}
   \caption[example] 
   { \label{fig:e0102_s3_n0016_1} 
Fitted values from the E0102 S3 data with the N0016 contamination model. LEFT: global normalization vs. time.  RIGHT: \ion{O}{vii}~He$\alpha$~{\em r} line normalization in units of ${\mathrm{photons~cm^{-2}~s^{-1}}}$ vs. time. The solid black horizontal line indicates the IACHEC value and the dotted lines indicate $\pm10\%$ from the IACHEC value.}
   \end{figure}

   \begin{figure} [ht]
   \begin{center}
  \includegraphics[trim={20 150 10 180}, clip, height=6.7cm]{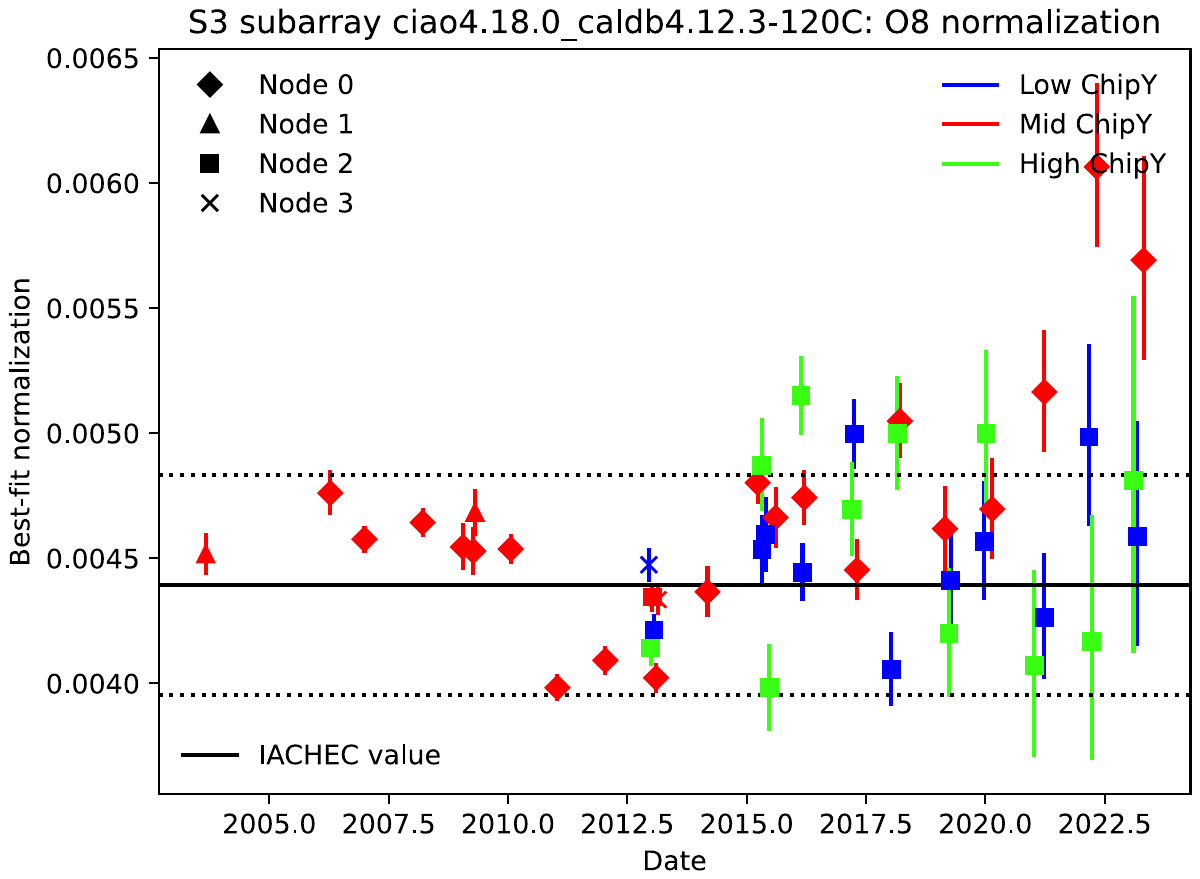}
  \includegraphics[trim={20 150 10 180}, clip, height=6.7cm]{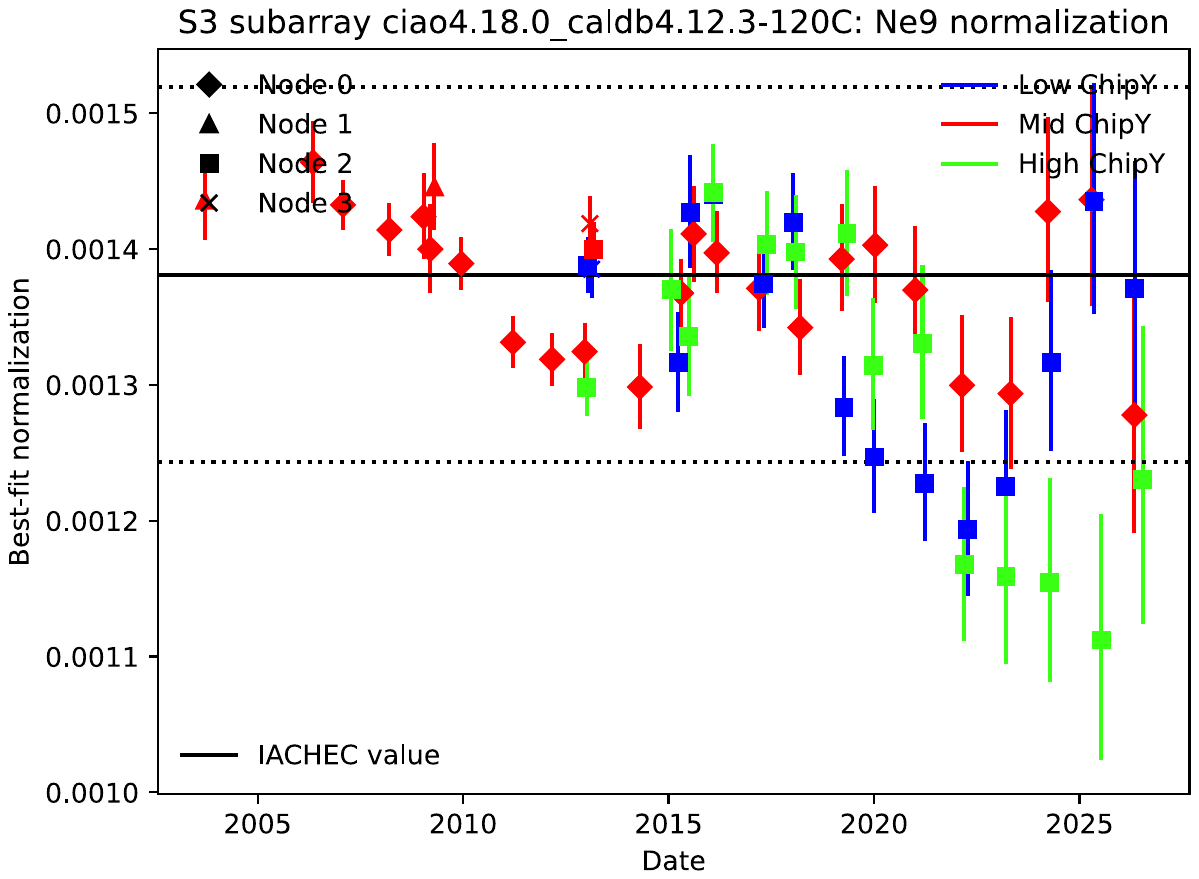}
   \end{center}
   \vspace{-0.8cm}
   \caption[example] 
   { \label{fig:e0102_s3_n0016_2} 
Fitted values from the E0102 S3 data with the N0016 contamination model. LEFT:  \ion{O}{viii}~Ly$\alpha$ line normalization in units of ${\mathrm{photons~cm^{-2}~s^{-1}}}$  vs. time.  RIGHT: \ion{Ne}{ix}~He$\alpha$~{\em r} line normalization in units of ${\mathrm{photons~cm^{-2}~s^{-1}}}$ vs. time. The solid black horizontal line indicates the IACHEC value and the dotted lines indicate $\pm10\%$ from the IACHEC value. }
   \end{figure}

     \begin{figure} [ht]
   \begin{center}
  \includegraphics[trim={20 150 10 180}, clip, height=6.7cm]{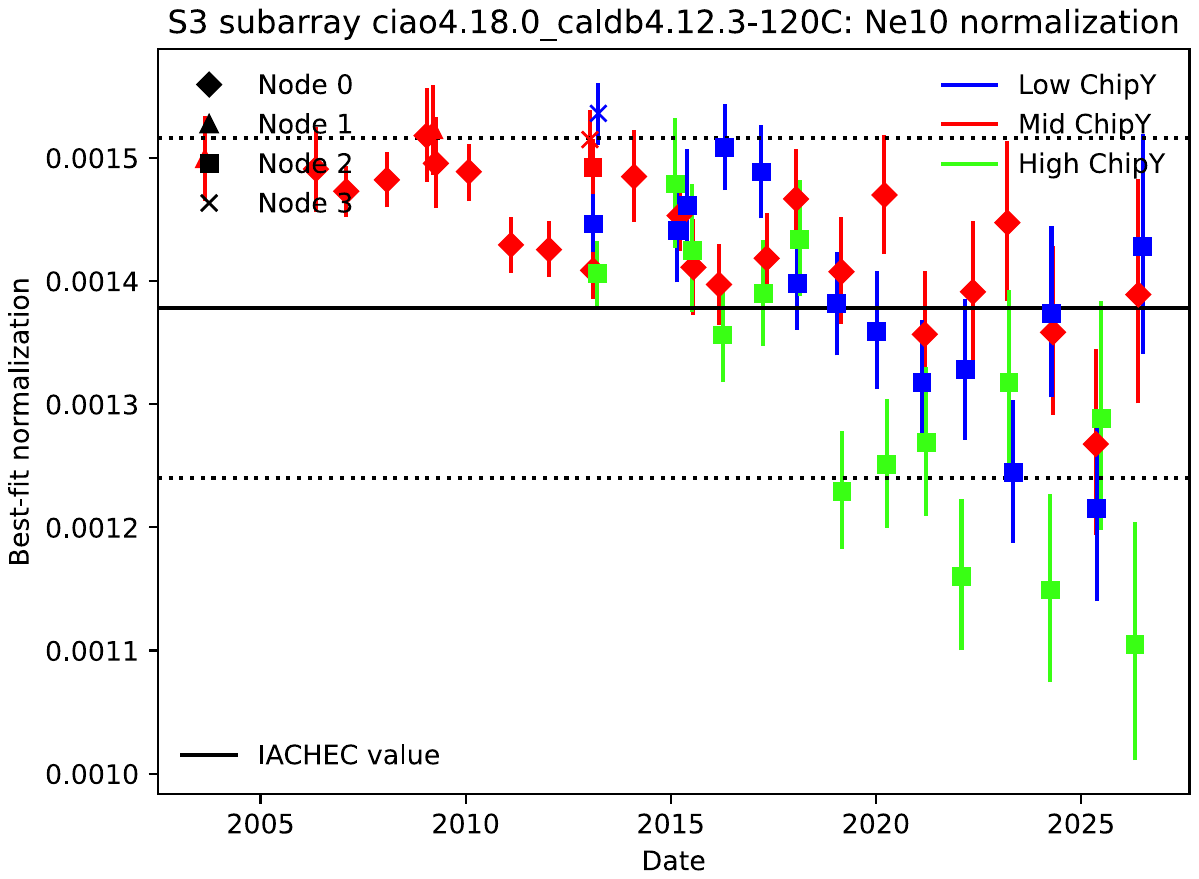}
  \includegraphics[trim={20 150 10 180}, clip, height=6.7cm]{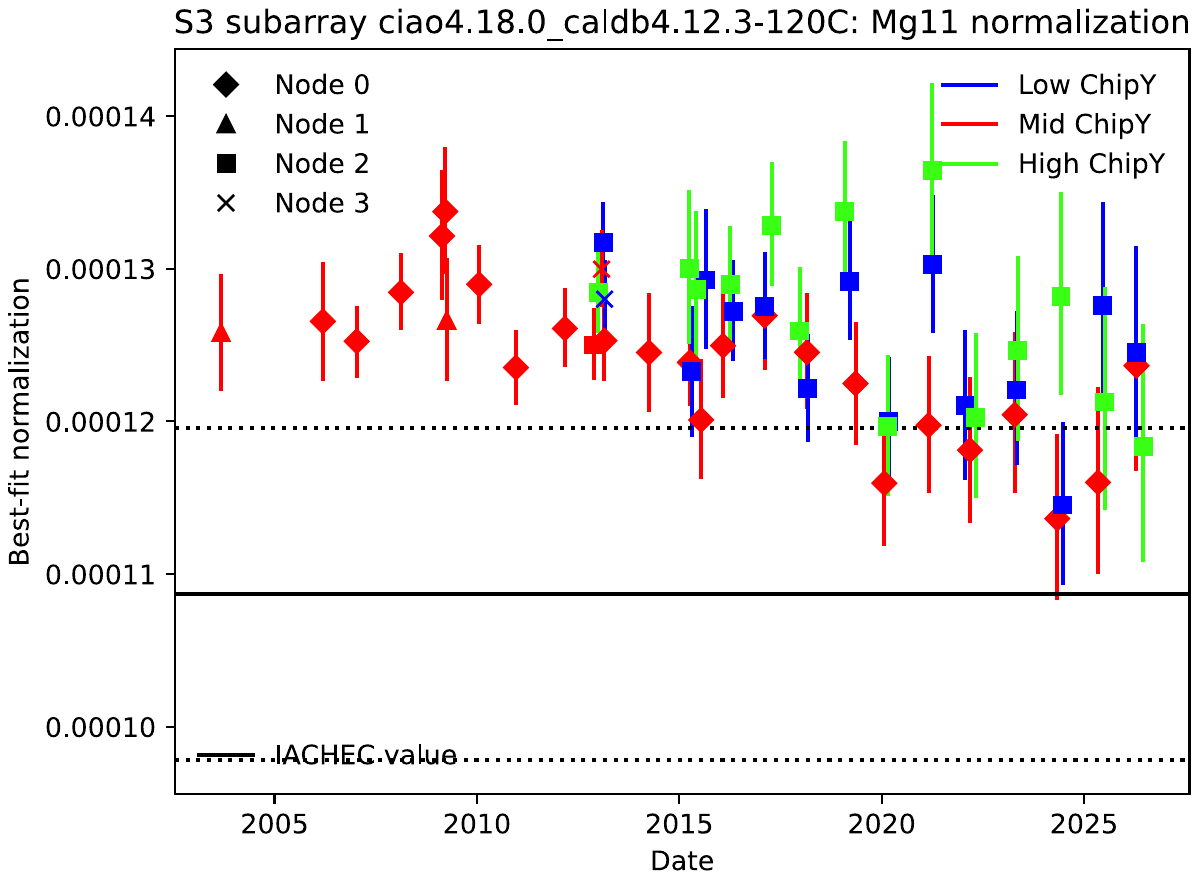}
   \end{center}
   \vspace{-0.8cm}
   \caption[example] 
   { \label{fig:e0102_s3_n0016_3} 
Fitted values from the E0102 S3 data with the N0016 contamination model. LEFT: \ion{Ne}{X}~Ly$\alpha$ line normalization in units of ${\mathrm{photons~cm^{-2}~s^{-1}}}$  vs. time.  RIGHT: \ion{Mg}{XI}~He$\alpha$~{\em r} line normalization in units of ${\mathrm{photons~cm^{-2}~s^{-1}}}$ vs. time. The solid black horizontal line indicates the IACHEC value and the dotted lines indicate $\pm10\%$ from the IACHEC value. }
   \end{figure}

\begin{longtable}{rcrrcrrcrc} 
\caption{ACIS I3 Observations of E0102} 
\label{tab:i3obs} \\

\toprule
ObsID   & Date     & ChipX  & ChipY     & Node  & Exposure      & Counts        & Frame & 1stRow & Nrows \\
        &          &        &           &       & (s)           & (0.3-2keV)    & (s)   &         &  \\
\midrule
\endfirsthead

\caption[]{ACIS I3 Observations of E0102 -- continued} \\

\toprule
ObsID   & Date     & ChipX  & ChipY     & Node  & Exposure      & Counts        & Frame & 1stRow & Nrows \\
        &          &        &           &       & (s)           & (0.3-2keV)    & (s)   &         &  \\
\midrule
\endhead

\bottomrule
\endfoot

 3526	& 2003.59  & 884.6  & 120.8	& 3	 & 7859.9	 & 21368	& 1.0	& 1	& 256  \\	
 6756	& 2006.20  & 888.7  & 922.5	& 3	 & 7159.9	 & 23197	& 0.8	& 768	& 256	\\
 9690	& 2008.09  & 884.4  & 925.6	& 3	 & 19192.5	 & 60496	& 0.8	& 768	& 256	\\
10649	& 2009.05  & 880.3  & 926.4	& 3	 & 7637.5	 & 23510	& 0.8	& 768	& 256	\\
11956	& 2009.98  & 882.6  & 922.3	& 3	 & 19189.4	 & 57787	& 0.8	& 768	& 256	\\
13092	& 2011.08  & 875.1  & 940.0	& 3	 & 19050.1	 & 53984	& 0.8	& 768	& 256	\\
14257	& 2012.04  & 881.9  & 928.7	& 3	 & 19051.0	 & 51539	& 0.8	& 768	& 256	\\
15466	& 2013.08  & 875.5  & 937.8	& 3	 & 19082.9	 & 47984	& 0.8	& 768	& 256	\\
15471	& 2013.09  & 877.4  & 505.8	& 3	 & 9574.3	 & 22301	& 0.8	& 341	& 256	\\
15472	& 2013.09  & 873.9  & 144.6	& 3	 & 9574.3	 & 16405	& 0.8	&  1	& 256	\\
16588	& 2014.24  & 875.4  & 953.9	& 3	 & 9571.9	 & 21634	& 0.8	& 768	& 256	\\
17379	& 2015.15  & 872.9  & 929.4	& 3	 & 17616.6	 & 35717	& 0.8	& 768	& 256	\\
17687	& 2015.54  & 866.5  & 934.0	& 3	 & 13285.4	 & 25598	& 0.8	& 768	& 256	\\
18417	& 2016.20  & 870.5  & 938.8	& 3	 & 22886.1	 & 40565	& 0.8	& 768	& 256	\\
19849   & 2017.21  & 877.0  & 938.6     & 3      & 23837.4       & 36761        & 0.8   & 768   & 256   \\  
20638   & 2018.20  & 877.7  & 933.7     & 3      & 23838.1       & 32164        & 0.8   & 768   & 256   \\
21803   & 2019.12  & 872.6  & 938.2     & 3      & 21935.7       & 25694        & 0.8   & 768   & 256   \\
21807   & 2019.18  & 876.2  & 506.5     & 3      & 26694.1       & 28275        & 0.8   & 341   & 256   \\
21808   & 2019.18  & 865.2  & 146.3     & 3      & 33354.0       & 23527        & 0.8   &   1   & 256   \\
22809   & 2020.09  & 863.3  & 151.6     & 3      & 33353.2       & 20416        & 0.8   &   1   & 256   \\
22804   & 2020.09  & 872.4  & 941.0     & 3      & 23838.2       & 24039        & 0.8   & 768   & 256   \\
22808   & 2020.12  & 877.6  & 504.5     & 3      & 26695.0       & 24572        & 0.8   & 341   & 256   \\
24576   & 2021.09  & 877.1  & 933.9     & 3      & 21934.90      & 18831        & 0.8   & 768   & 256   \\
24580   & 2021.17  & 878.9  & 505.8     & 3      & 26694.81      & 20495        & 0.8   & 341   & 256   \\
24581   & 2021.19  & 869.9  & 147.7     & 3      & 30499.63      & 14962        & 0.8   &   1   & 256   \\
25617   & 2022.23  & 877.2  & 942.2     & 3      & 24847.78      & 17847        & 0.8   & 768   & 256   \\
25621   & 2022.19  & 879.0  & 511.1     & 3      & 11476.68      & 7689         & 0.8   & 341   & 256   \\
26358   & 2022.20  & 880.0  & 509.5     & 3      & 17012.64      & 11052        & 0.8   & 341   & 256   \\
25622   & 2022.19  & 869.3  & 152.2     & 3      & 16234.17      & 6821         & 0.8   &   1   & 256   \\
26359   & 2022.19  & 870.4  & 152.6     & 3      & 17182.22      & 7518         & 0.8   &   1   & 256   \\
26986   & 2023.20  & 880.8  & 934.2     & 3      & 12901.45      & 7817         & 0.8   & 768   & 256   \\
27745   & 2023.20  & 881.8  & 933.5     & 3      & 12329.51      & 7550         & 0.8   & 768   & 256   \\
26990   & 2023.23  & 884.5  & 497.7     & 3      & 17659.02      & 9813         & 0.8   & 341   & 256   \\
27762   & 2023.24  & 883.9  & 501.0     & 3      & 9575.09       & 5278         & 0.8   & 341   & 256   \\
26991   & 2023.24  & 874.4  & 140.8     & 3      & 16232.63      & 5743         & 0.8   &   1   & 256   \\
27773   & 2023.24  & 875.7  & 141.1     & 3      & 15280.66      & 5431         & 0.8   &   1   & 256   \\
28436   & 2024.33  & 870.3  & 150.6     & 3      & 33354.77      & 9980         & 0.8   &   1   & 256   \\
28435   & 2024.35  & 879.6  & 505.3     & 3      & 27646.10      & 12596        & 0.8   & 341   & 256   \\
28431   & 2024.38  & 875.3  & 944.2     & 3      & 23839.71      & 11928        & 0.8   & 768   & 256   \\
29584   & 2025.36  & 875.0  & 947.1     & 3      & 23838.93      & 9849         & 0.8   & 768   & 256   \\
29589   & 2025.36  & 869.8  & 158.9     & 3      & 33353.42      & 8137         & 0.8   &   1   & 256   \\
29588   & 2025.36  & 878.8  & 514.3     & 3      & 28596.52      & 10818        & 0.8   & 341   & 256   \\
31376   & 2026.35  & 876.5  & 521.3     & 3      & 28598.88      & 9222         & 0.8   & 341   & 256   \\
31372   & 2026.38  & 874.0  & 954.8     & 3      & 23839.79      & 7906         & 0.8   & 768   & 256   \\
31377   & 2026.39  & 866.5  & 163.0     & 3      & 33353.22      & 7055         & 0.8   &   1   & 256   \\
\end{longtable}

   \begin{figure} [ht]
   \begin{center}
  \includegraphics[trim={20 150 10 180}, clip, height=6.7cm]{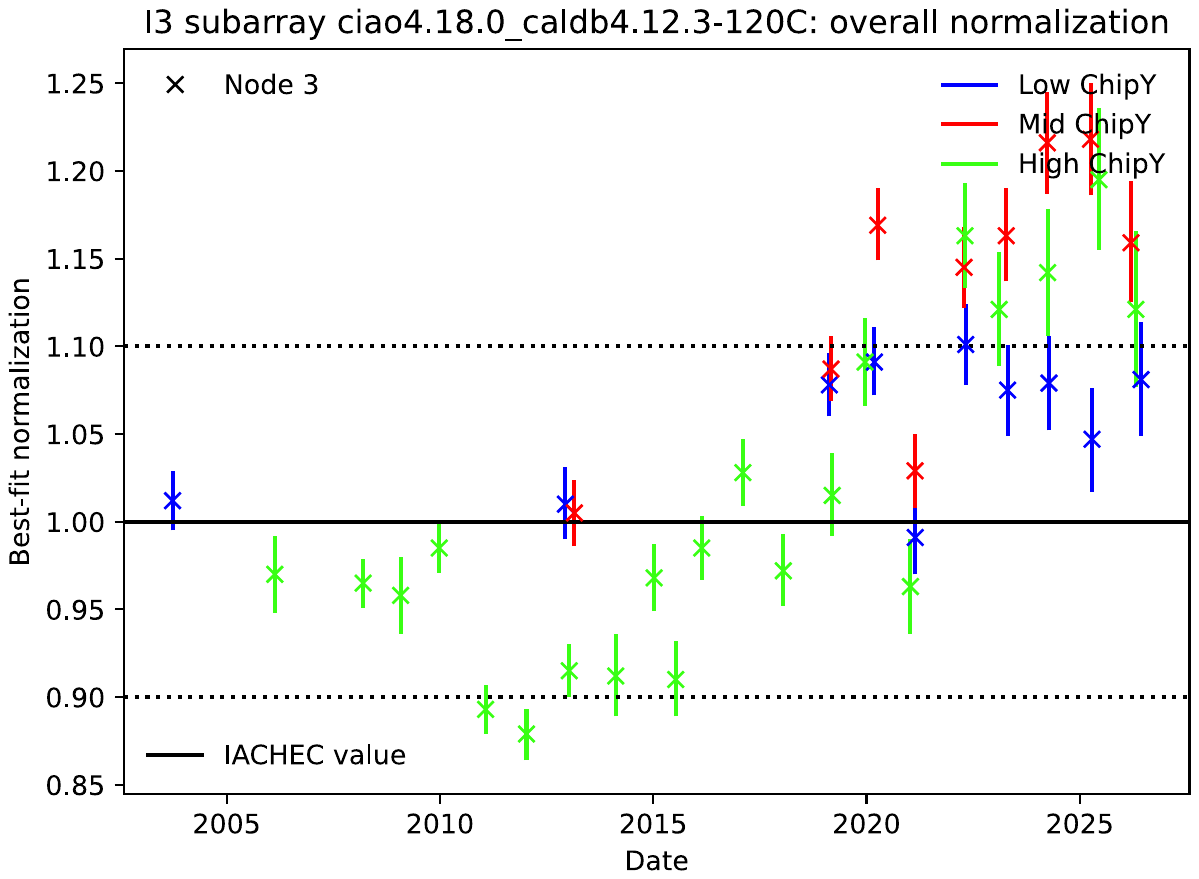}
  \includegraphics[trim={20 150 10 180}, clip, height=6.7cm]{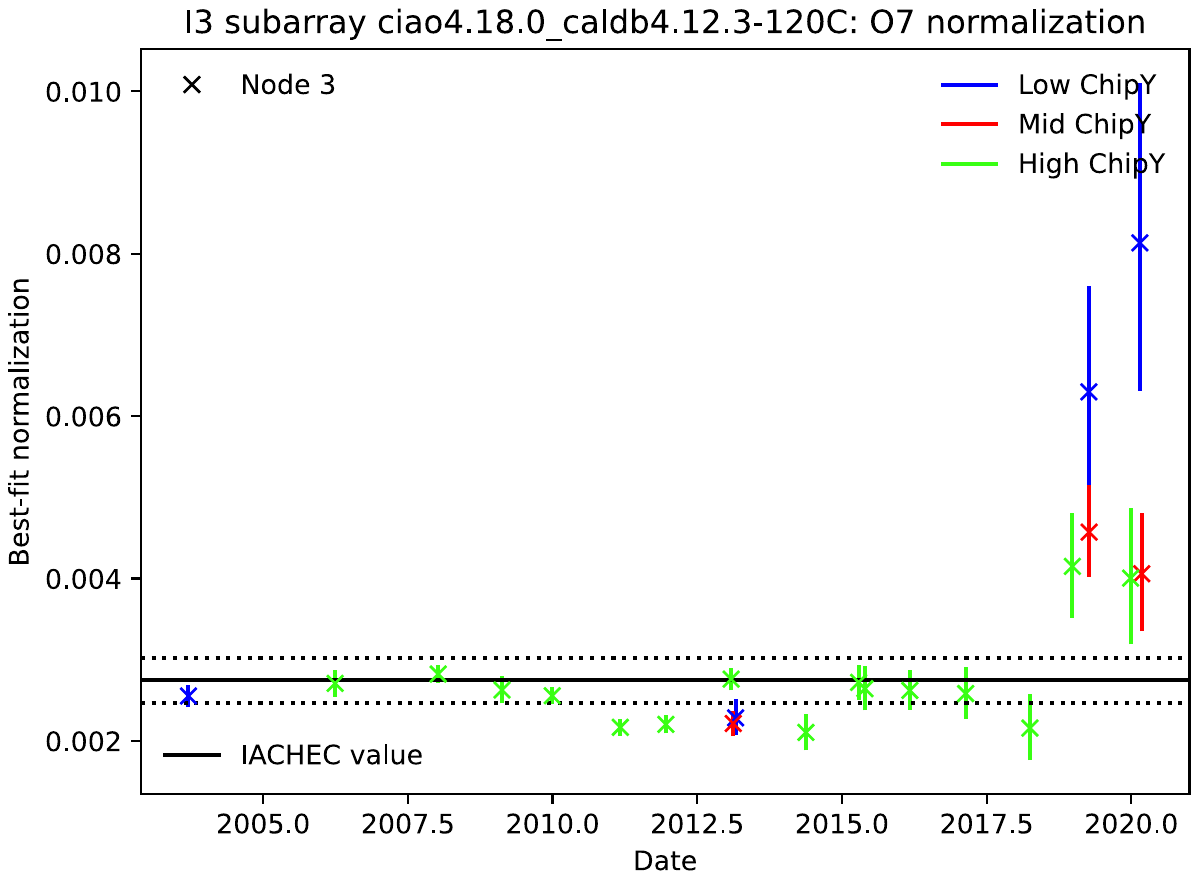}
   \end{center}
   \caption[example] 
   { \label{fig:e0102_i3_n0016_1} 
Fitted values from the E0102 I3 data with the N0016 contamination model. LEFT: global normalization vs. time.  RIGHT: \ion{O}{vii}~He$\alpha$~{\em r} line normalization in units of ${\mathrm{photons~cm^{-2}~s^{-1}}}$ vs. time. The solid black horizontal line indicates the IACHEC value and the dotted lines indicate $\pm10\%$ from the IACHEC value.
}
   \end{figure}

   \begin{figure} [ht]
   \begin{center}
  \includegraphics[trim={20 150 10 180}, clip, height=6.7cm]{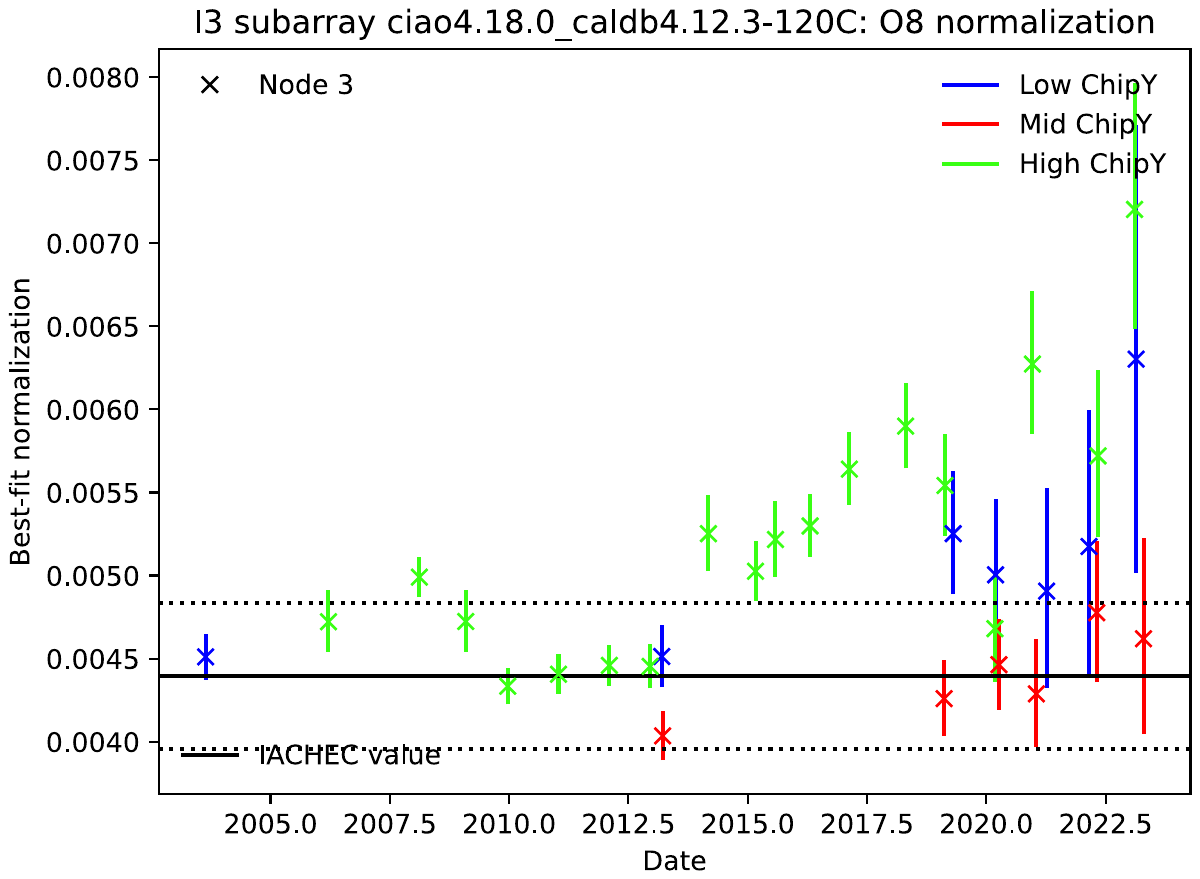}
  \includegraphics[trim={20 150 10 180}, clip, height=6.7cm]{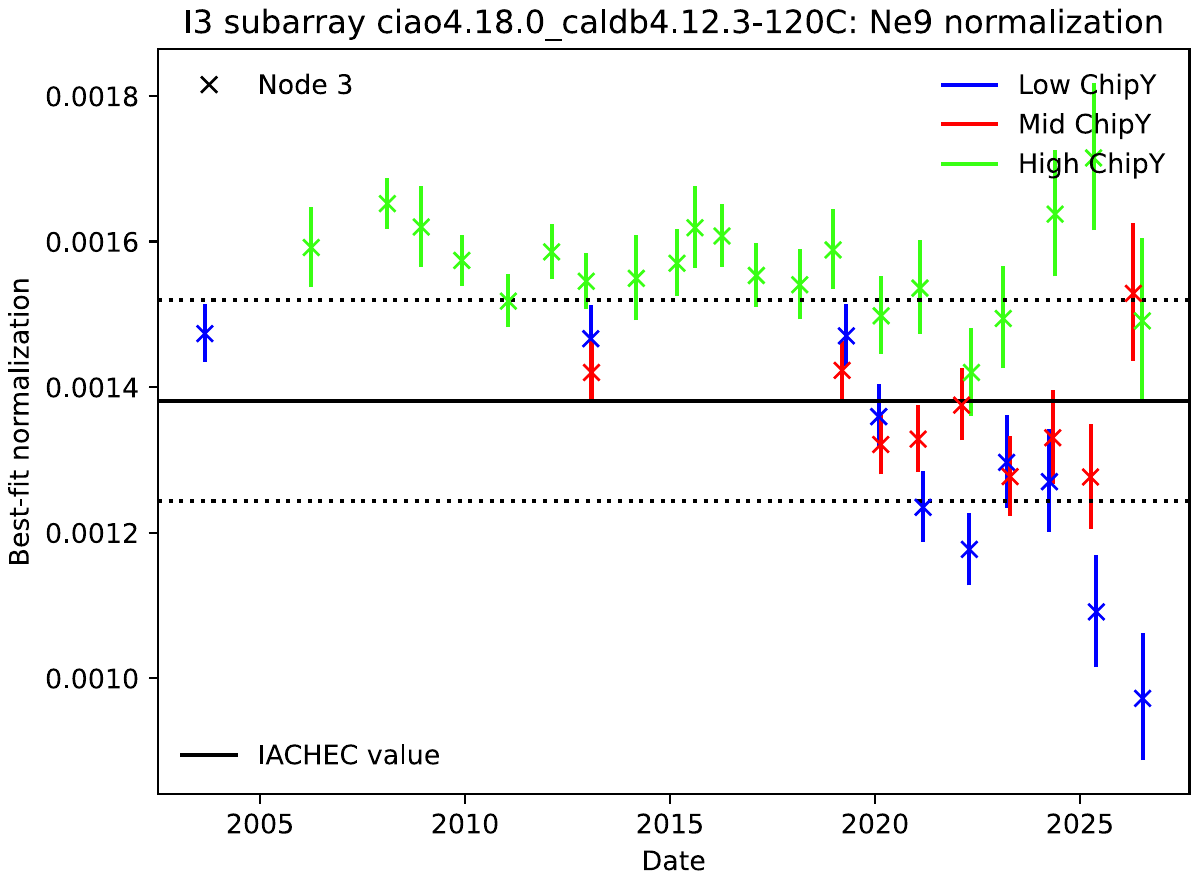}
   \end{center}
   \caption[example] 
   { \label{fig:e0102_i3_n0016_2} 
Fitted values from the E0102 I3 data with the N0016 contamination model. LEFT:  \ion{O}{viii}~Ly$\alpha$ line normalization in units of ${\mathrm{photons~cm^{-2}~s^{-1}}}$  vs. time.  RIGHT: \ion{Ne}{ix}~He$\alpha$~{\em r} line normalization in units of ${\mathrm{photons~cm^{-2}~s^{-1}}}$ vs. time. The solid black horizontal line indicates the IACHEC value and the dotted lines indicate $\pm10\%$ from the IACHEC value.}
   \end{figure} 

  \begin{figure} [ht]
   \begin{center}
  \includegraphics[trim={20 150 10 180}, clip, height=6.7cm]{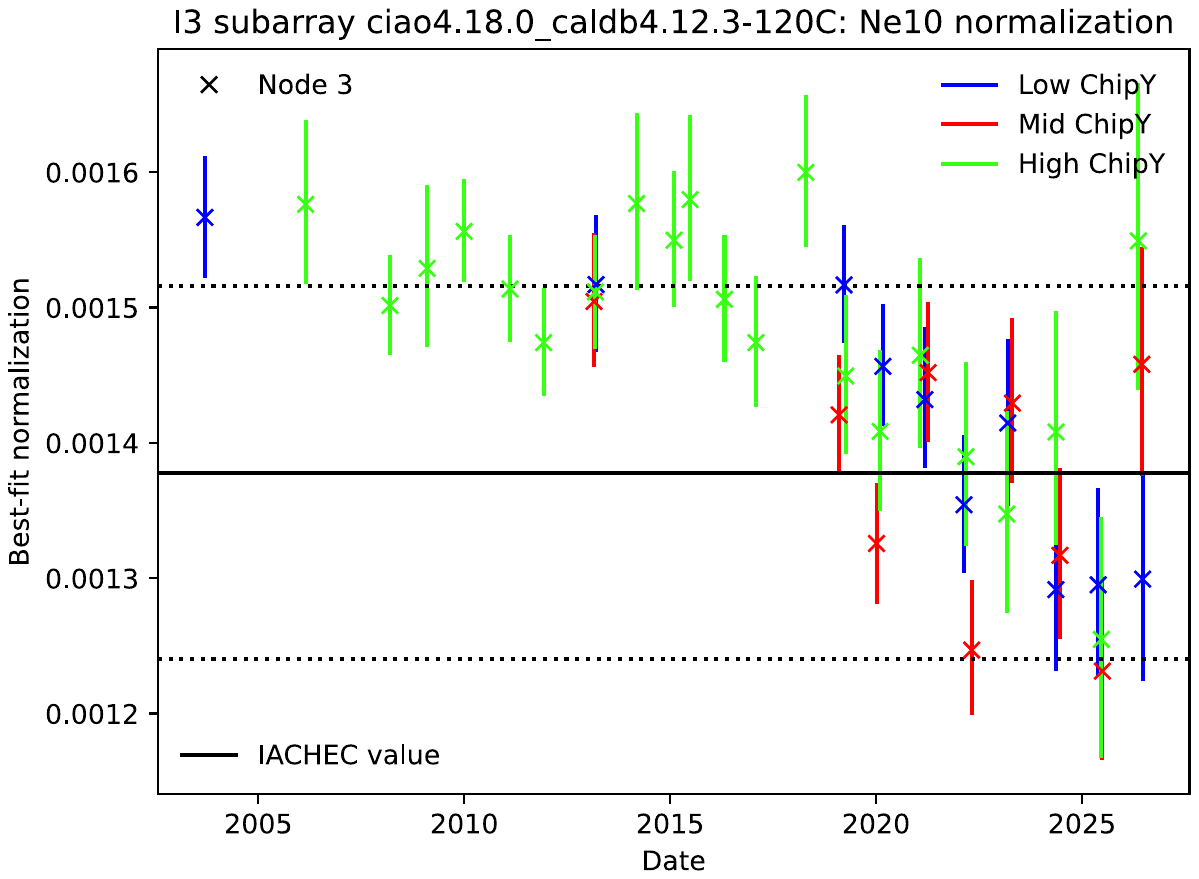}
  \includegraphics[trim={20 150 10 180}, clip, height=6.7cm]{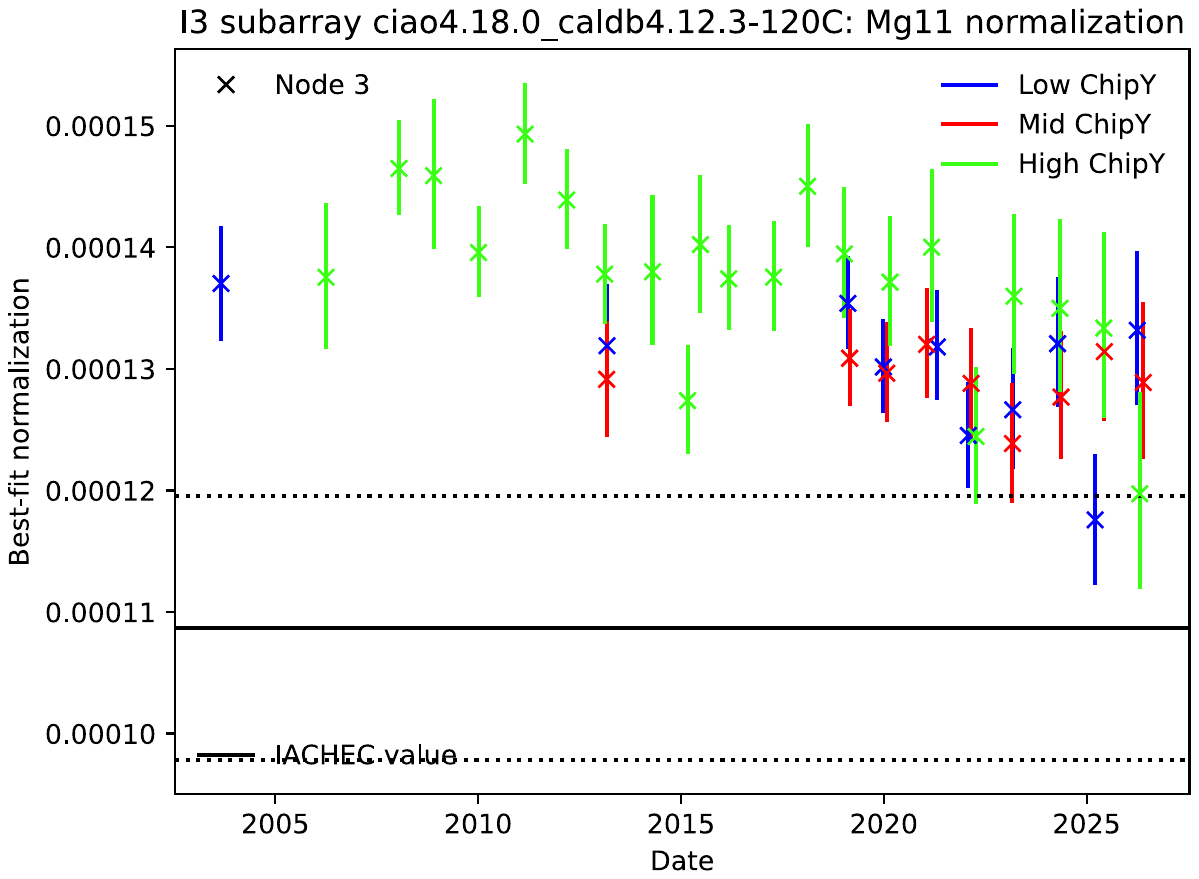}
   \end{center}
   \caption[example] 
   { \label{fig:e0102_i3_n0016_3} 
Fitted values from the E0102 I3 data with the N0016 contamination model. LEFT: \ion{Ne}{X}~Ly$\alpha$ line normalization in units of ${\mathrm{photons~cm^{-2}~s^{-1}}}$  vs. time.  RIGHT: \ion{Mg}{XI}~He$\alpha$~{\em r} line normalization in units of ${\mathrm{photons~cm^{-2}~s^{-1}}}$ vs. time. The solid black horizontal line indicates the IACHEC value and the dotted lines indicate $\pm10\%$ from the IACHEC value.}
   \end{figure}

The results for the S3 CCD in ACIS (a back-illuminated detector) are shown in Figure~\ref{fig:e0102_s3_n0016_1} (global normalization and \ion{O}{vii}~He$\alpha$~{\em r}), in Figure~\ref{fig:e0102_s3_n0016_2} (\ion{O}{viii}~Ly$\alpha$ and \ion{Ne}{ix}~He$\alpha$~{\em r}), and in Figure~\ref{fig:e0102_s3_n0016_3} (\ion{Ne}{X}~Ly$\alpha$ and \ion{Mg}{xi}~He$\alpha$~{\em r}). On the S3 CCD, the aimpoint of the CXO telescope is close to the mid chipy position. 
The global normalization is consistent with the IACHEC value within 10\% from 2003 until 2015, but tends to be $\sim7\%$ higher.  After 2015, the global normalization is 10-15 \% higher than the IACHEC value with the high chipy value being up to 20\% higher.
The \ion{O}{vii}~He$\alpha$~{\em r} line normalizations are consistent with the IACHEC value within $\pm10\%$ until 2013.  After 2015 the line normalizations are much higher than the IACHEC value and the uncertainties become so large the data are not useful after 2020.
The \ion{O}{viii}~Ly$\alpha$ line normalizations are consistent within the uncertainties through 2020 but begin to disagree in 2021 with the mid chipy positions disagreeing the most.  By 2022 the statistical uncertainties had increased such that the measurements are not useful to measure the additional absorption. The \ion{Ne}{ix}~He$\alpha$~{\em r} line normalizations at the mid chipy position are consistent within $10\%$ of the IACHEC value through 2026.  The low and high chipy positions show larger disagreement with the IACHEC value, with the high chipy position being as much as 
$15\%$ lower than the IACHEC value in recent years. The \ion{Ne}{X}~Ly$\alpha$ line normalizations at the mid chipy position are consistent with each other and the IACHEC value to within $10\%$ for all measurements except for the 2024 measurement. The low and high chipy measurements show more scatter with time and the high chipy measurements appear to be trending downwards with time, disagreeing with the IACHEC value by as much as 
$20\%$.  It is interesting to note that the high chipy measurements for both \ion{Ne}{ix}~He$\alpha$~{\em r}  and \ion{Ne}{X}~Ly$\alpha$ have a downward trend in time from 2020 to 2026.  This could be explained as a deficiency in the spatial model for the contamination layer or a change in the detector response due to increasing charge transfer inefficiency (CTI) or a combination of both effects.  The \ion{Mg}{xi}~He$\alpha$~{\em r}
line normalizations are consistent with each other to within $15\%$ for all positions but are systematically higher than the IACHEC value by $\sim15\%$. Other instrument such as the CCDs on the {\em {Suzaku}} and {\em{eROSITA}} missions show a similar discrepancy with the IACHEC standard value for the \ion{Mg}{xi}~He$\alpha$~{\em r} normalization.  The \ion{Mg}{xi}~He$\alpha$~{\em r} normalization did not receive much attention from the IACHEC thermal supernova remnants working group when the standard model was developed early in the CXO and {\em{XMM-Newton}} missions and warrants further scrutiny.

The results for the I3 CCD in ACIS (a front-illuminated detector) are shown in Figure~\ref{fig:e0102_i3_n0016_1} (global normalization and \ion{O}{vii}~He$\alpha$~{\em r}), in Figure~\ref{fig:e0102_i3_n0016_2} (\ion{O}{viii}~Ly$\alpha$ and \ion{Ne}{ix}~He$\alpha$~{\em r}), and in Figure~\ref{fig:e0102_i3_n0016_3} (\ion{Ne}{X}~Ly$\alpha$ and \ion{Mg}{xi}~He$\alpha$~{\em r}). On the I3 CCD, the aimpoint is close to the high chipy position.   The global normalizations for I3 are consistent with each other and the IACHEC value to within $\pm12\%$ from 2003 to 2019.  After 2019, the global normalization is consistently higher than the IACHEC value with the mid chipy position being as much as $22\%$ higher.  The \ion{O}{vii}~He$\alpha$~{\em r} line normalizations are consistent with the IACHEC value within $\pm15\%$ until 2018.  After 2018 the line normalizations are much higher than the IACHEC value and the uncertainties become so large the data are not useful after 2020. 
The \ion{O}{viii}~Ly$\alpha$ line normalizations are consistent with each other and the IACHEC value to within $\pm12\%$ up to 2013.  After 2013 the scatter increases significantly and the high chipy measurements can disagree with the IACHEC value by $\sim30\%$. After 2023, we do not report results for the \ion{O}{viii}~Ly$\alpha$ line normalization given how large the statistical uncertainties are. 
The \ion{Ne}{ix}~He$\alpha$~{\em r} line normalizations at the high chipy position are consistent with each other to within $\pm10\%$ for the entire mission, but are consistently $\sim15\%$ higher than the IACHEC value.  This is somewhat surprising because the spectral resolution is lowest at high chipy on the I3 CCD.  The mid and low chipy normalizations are more consistent with the IACHEC value, but the low chipy values appear to be trending down from 2020 to 2026.  The most recent measurement in 2026 is $\sim25\%$ lower than the high and mid chipy positions and the IACHEC value.  It should be noted that the spectral resolution on I3 is the highest at low chipy as shown in Figure~\ref{fig:e0102_spectra_1}.
The \ion{Ne}{X}~Ly$\alpha$ line normalizations at the high chipy position are consistent with each other to within $10\%$ through 2019 but are systematially about $10\%$ higher than the IACHEC value.  After 2019, all positions are in better agreement with the IACHEC value although the scatter gets larger. There may be a downward trend in time for the high chipy measurements, but the most recent measurement in 2026 was considerably above the most recent values.  The \ion{Mg}{xi}~He$\alpha$~{\em r} line normalizations are consistent with each other to within $20\%$ for all positions but are systematically higher than the IACHEC value by $\sim20\%$.   This difference with respect to the IACHEC value is similar to the S3 results and that of other missions mentioned previously.

 \begin{figure} [ht]
   \begin{center}
  \includegraphics[trim={20 150 10 180}, clip, height=6.7cm]{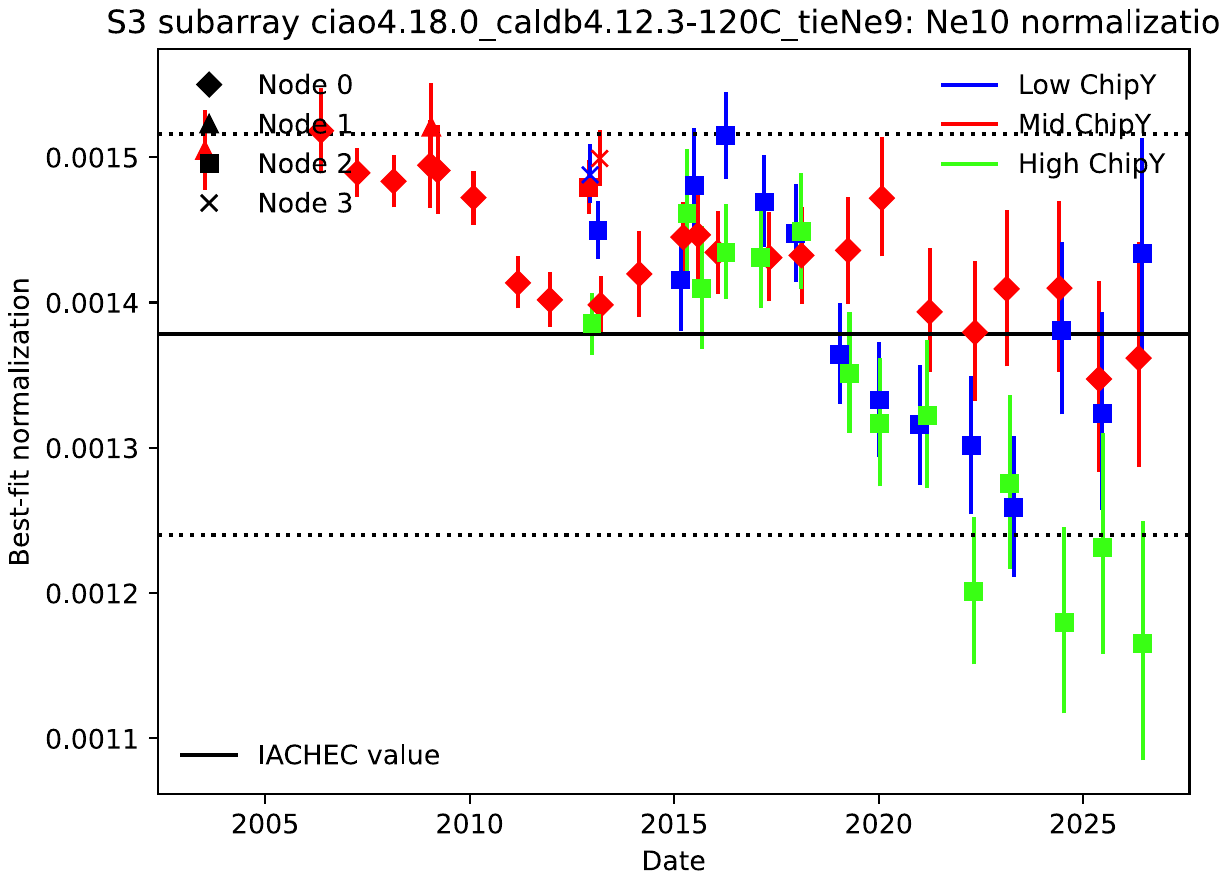}
  \includegraphics[trim={20 150 10 180}, clip, height=6.7cm]{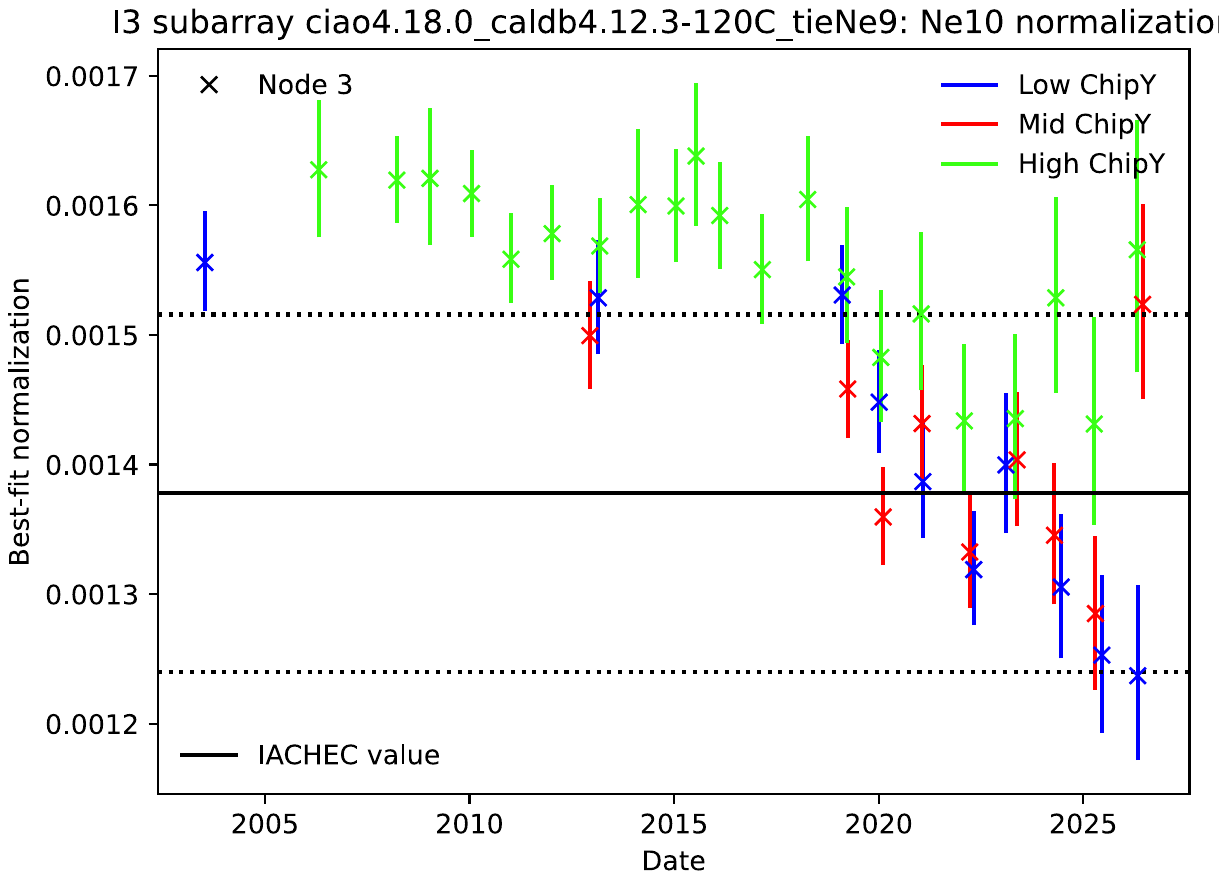}
   \end{center}
   \caption[example] 
   { \label{fig:e0102_ne10_tie_ne9} 
Fitted values for the \ion{Ne}{X}~Ly$\alpha$ line normalization with the \ion{Ne}{ix}~He$\alpha$~{\em r} line normalization fixed to the \ion{Ne}{X}~Ly$\alpha$ line normalization with the IACHEC model value. LEFT: S3 \ion{Ne}{X}~Ly$\alpha$ line normalization in units of ${\mathrm{photons~cm^{-2}~s^{-1}}}$  vs. time.  RIGHT: I3 \ion{Mg}{XI}~He$\alpha$~{\em r} line normalization in units of ${\mathrm{photons~cm^{-2}~s^{-1}}}$ vs. time. The solid black horizontal line indicates the IACHEC value and the dotted lines indicate $\pm10\%$ from the IACHEC value.}
   \end{figure}

We repeated this analysis with the \ion{Ne}{ix}~He$\alpha$~{\em r} line normalization set to the \ion{Ne}{X}~Ly$\alpha$ line normalization by the ratio of these two normalizations determined by the fits from 2003 to 2010 data.  During the fitting process, the \ion{Ne}{X}~Ly$\alpha$ line normalization was allowed to vary while the \ion{Ne}{ix}~He$\alpha$~{\em r} line normalization was not. The ratio we determined from fits to the 2003 to 2010 data is \ion{Ne}{ix}~He$\alpha$~{\em r} / \ion{Ne}{X}~Ly$\alpha$ = 0.95, which compares to the ratio in the IACHEC model of $\sim1.0$. The rationale for this action is that detector response issues such as CTI and gain may be affecting the distribution of counts between these two line complexes and the result might be more robust if the flux in a broader band was measured.  In particular, there may be detector response effects that are a function of position on the detector that may complicate our interpretation of the results with regards to the spatial model of the contamination layer.  The results for the \ion{Ne}{X}~Ly$\alpha$ line normalizations with this fitting approach are shown in Figure~\ref{fig:e0102_ne10_tie_ne9} for the S3 and I3 CCDs.  The results are similar to those shown in Figures~\ref{fig:e0102_s3_n0016_3} and ~\ref{fig:e0102_i3_n0016_3} when both line normalizations are fitted independently but there are small differences.  The scatter among the results for a specific detector position is reduced, for example, the scatter among the mid chipy measurements on S3 is smaller.  The low chipy and mid chipy values for the S3 CCD are more consistent with each other and more consistent with the IACHEC value, although there is still significant scatter in the low chipy values.
The downward trend for the S3 CCD at high chipy appears more significant in  Figure~\ref{fig:e0102_ne10_tie_ne9}.  This might be due to an inaccuracy in the spatial model of the contamination layer or the larger impact of CTI at higher chipy values or both.
The high chipy values for the I3 CCD are more consistent with each other and more consistent with the IACHEC value.  It appears that this approach is producing results with less scatter for the high chipy position on I3 where the spectral resolution is the lowest.
The mid chipy results had a downard trend from 2020 to 2025 but the 2026 point is significantly higher.  This variation in time is difficult to explain. The low chipy measurements have a clear downward trend from 2020 to 2026.  This could be explained by an inaccuracy in the spatial model of the contamination layer but would be difficult to explain as a detector response issue since the spectral resolution and gain are calibrated the best at the low chipy position on I3.  We note that the CXO experienced its lowest perigee altitude of the mission in 2023 during which the spacecraft transited a region of the radiation belts for the first time in the mission.  It is possible that the particle population is different in this part of the radiation belts and produced a different type of radiation damage than earlier in the mission.  In the future, we intend to perform these fits with focal plane temperature dependent response products and to explore the accuracy of the CTI correction to determine their impact on the fitted results.

\section{EVALUATION OF THE N0016 MODEL WITH A1795}
\label{sec:a1795}  

 \begin{figure} [ht]
   \begin{center}
   \includegraphics[height=6cm]{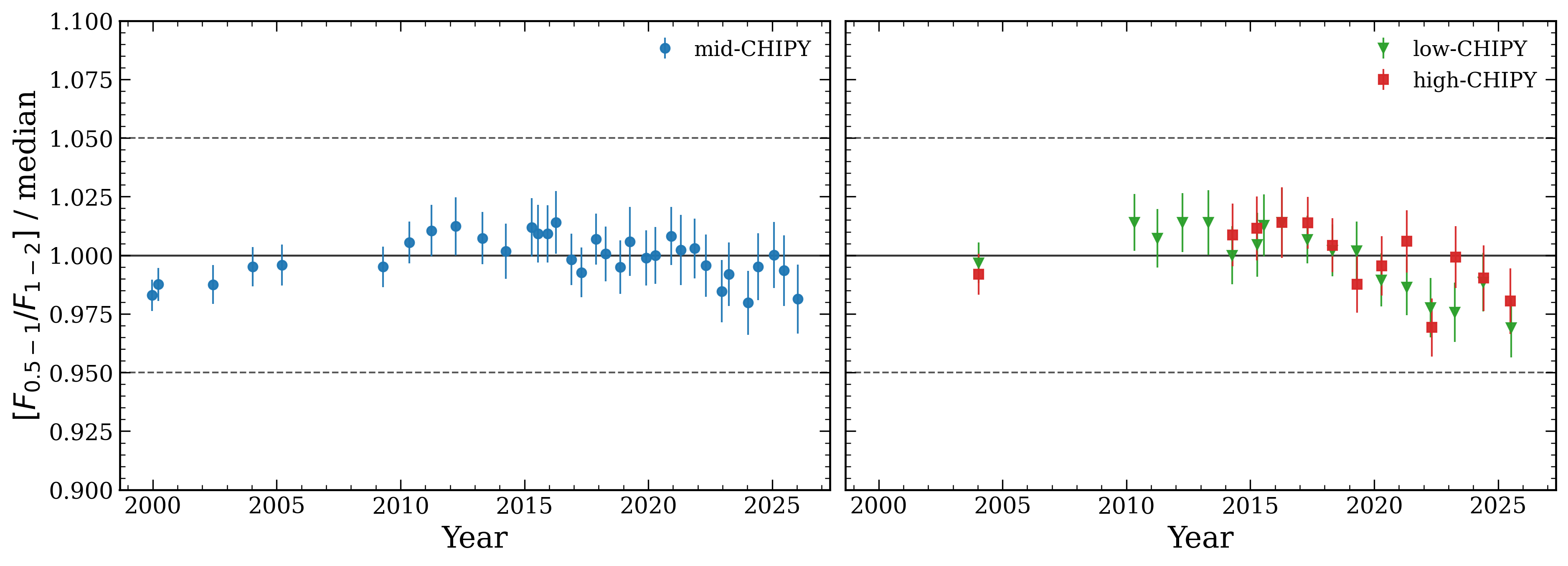}
   \end{center}
   \caption[example] 
   { \label{fig:a1795_n0016} 
Ratio of the $0.5-1.0$~keV to $1.0-2.0$~keV fluxes from the A1795 data on the S3 CCD with the N0016 contamination model. The ratios are normalized by the median value. The solid black line indicates the median, and the dashed lines show $\pm 5\%$ deviations from the median. \textit{Left}: mid-CHIPY observations. \textit{Right}: high/low-CHIPY observations.}
   \end{figure}

The A1795 data were analyzed to measure the relative stability of the $0.5-1.0$~keV and $1.0-2.0$~keV bands at different locations on the S3 CCD as a function of time. Since A1795 is a constant source like E0102, the ratio of these fluxes should remain stable if the contamination model correctly accounts for the time-dependent accumulation on the OBF. Using the flux ratio is particularly useful because it removes uncertainties associated with aperture placement, source normalization, and broad band calibration, while retaining sensitivity to residual errors in the contamination correction. This spectral analysis therefore provides a complementary test to the E0102 analysis. Indeed, the continuum is stronger in A1795, and the larger angular extent of the source samples a broader region of the OBF and detector. The bright central region of A1795 is used for this analysis.  Figure~\ref{fig:a1795_n0016} shows the ratio of the $0.5-1.0$~keV to $1.0-2.0$~keV fluxes, normalized by the median value, as a function of time. The left panel shows observations near the aimpoint, while the right panel shows observations near the bottom and top of the CCD, where the contaminant column is larger. The solid black line indicates the median ratio, and the dashed lines mark $\pm 5\%$ deviations from the median. The ratios are consistent over the full monitoring period, both near the aimpoint and at low/high CHIPY. This indicates that the N0016 contamination model preserves the relative soft band response of ACIS-S3 to within a few percent across the detector locations sampled by the A1795 observations.

   \begin{figure} [h]
   \begin{center}
   \includegraphics[height=10.0cm]{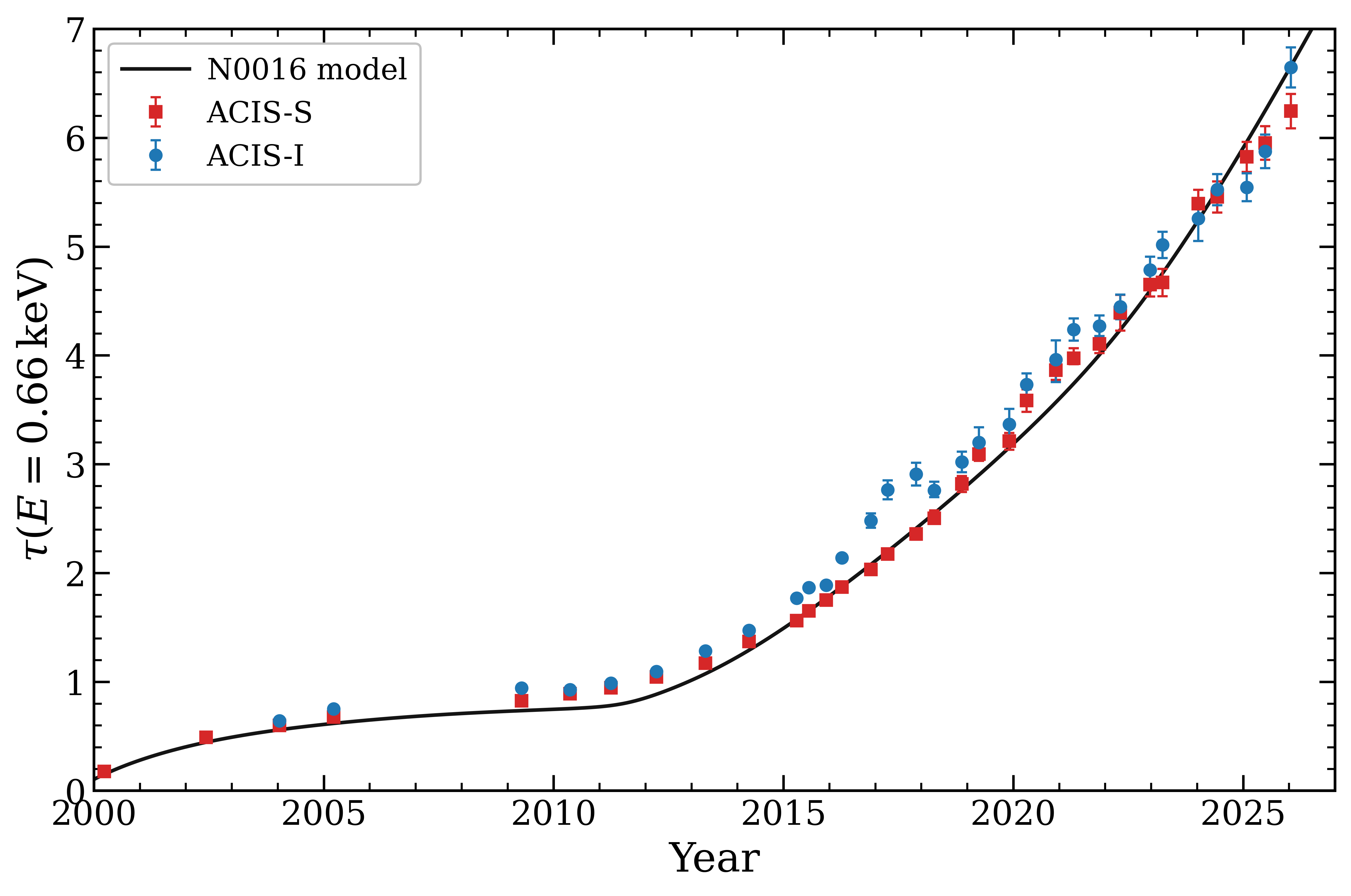}
   \end{center}
   \caption[example] 
   { \label{fig:a1795_optdepth} 
Optical depth of the ACIS contamination layer at 0.66~keV derived from A1795 observations near the nominal aimpoints on the S3 and I3 CCDs. Red squares show the S3 measurements, and blue circles show the I3 measurements. The solid black line shows the optical depth predicted by the N0016 contamination model.}
   \end{figure}  
   
The A1795 data were also used to measure the optical depth of the contamination layer at 0.66~keV near the aimpoint. The energy of 0.66~keV was selected because it is close to the Mn~L complex from the ACIS external calibration source (ECS), which provided an important early-mission measure of the low-energy contamination buildup. Figure~\ref{fig:a1795_optdepth} displays the optical depths computed from the A1795 data for the S3 and I3 CCDs. The data used in these plots were extracted close to the nominal aimpoints on the two CCDs. In general, the optical depths measured on S3 and I3 track each other well over the full time range, with only modest deviations between the two detectors. The black line indicates the optical depth predicted by the N0016 contamination model. Compared with the previous version of the model, the N0016 model provides a better description of the recent A1795 measurements and reproduces the continued increase in optical depth after 2020.

\begin{figure} [h]
   \begin{center}
   \includegraphics[height=10.0cm]{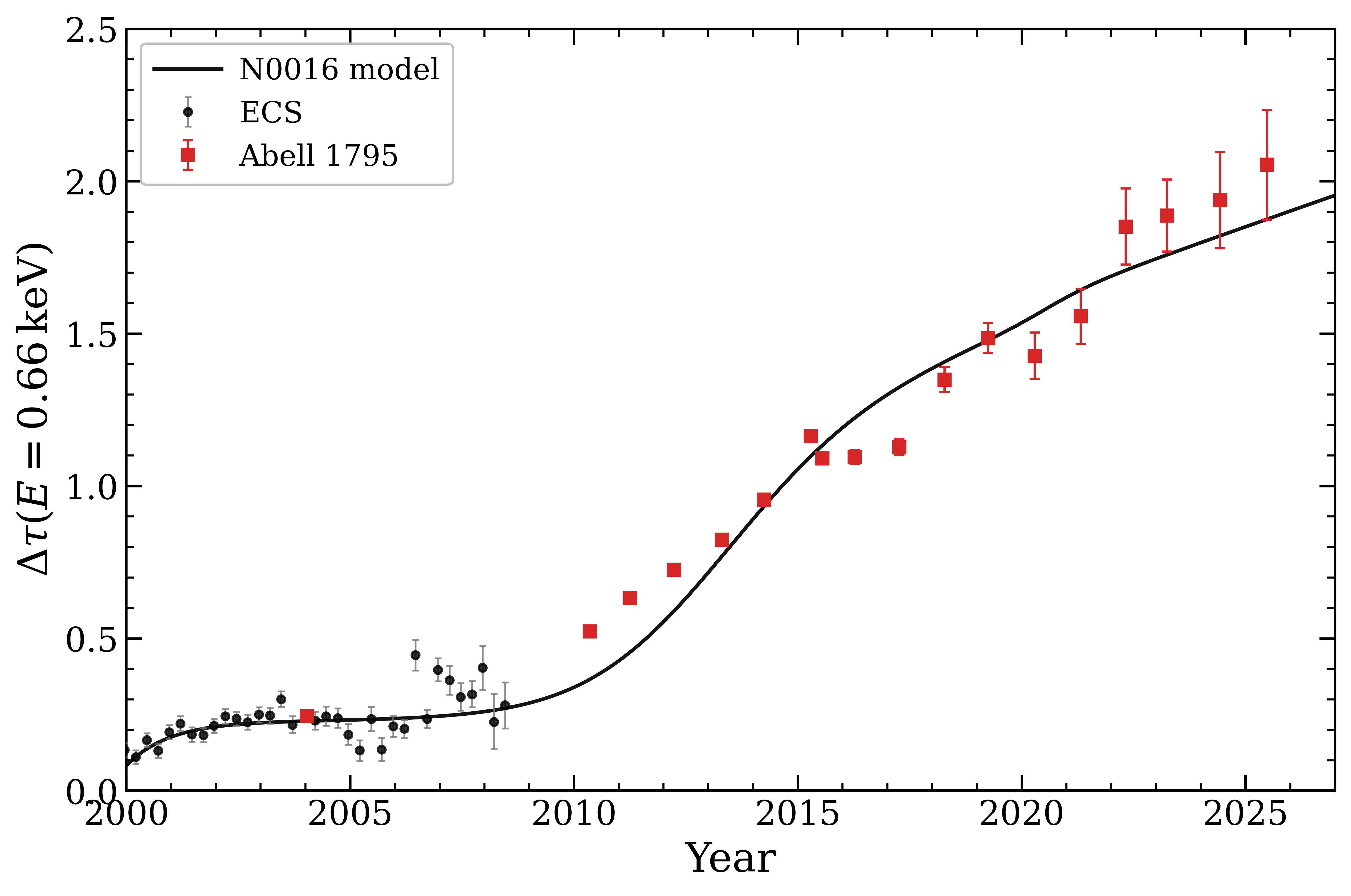}
   \end{center}
   \caption[example] 
   { \label{fig:a1795_optdepth_edge} 
Edge-to-center optical-depth difference at 0.66~keV on the S3 CCD. The measurements correspond to row 64, near the readout, relative to the center of the detector. Red squares show the A1795 measurements, and black circles show the ECS measurements. The solid black line shows the edge-to-center optical-depth difference predicted by the N0016 contamination model.}
   \end{figure}

The A1795 data are also used to measure the depth of the contaminant at the top and bottom of the ACIS S3 CCD relative to the center. The optical depths are derived at 0.66~keV and refer to the edge of the CCD at row 64, near the readout. Figure~\ref{fig:a1795_optdepth_edge} shows the evolution of this edge-to-center optical-depth difference using the ECS data at early times (black circles) and the A1795 data at later times (red squares). The optical depth difference was small early in the mission but has increased steadily since about 2009, indicating enhanced contaminant accumulation near the edge of the detector relative to the center. The N0016 contamination model captures this long term increase and provides a reasonable description of the A1795 measurements as a function of row number for ACIS S3.

\section{EVALUATION OF THE N0016 MODEL WITH Mrk~421}
\label{sec:Mrk421}  

\begin{figure} [ht]
   \begin{center}
   \includegraphics[height=6.0cm]{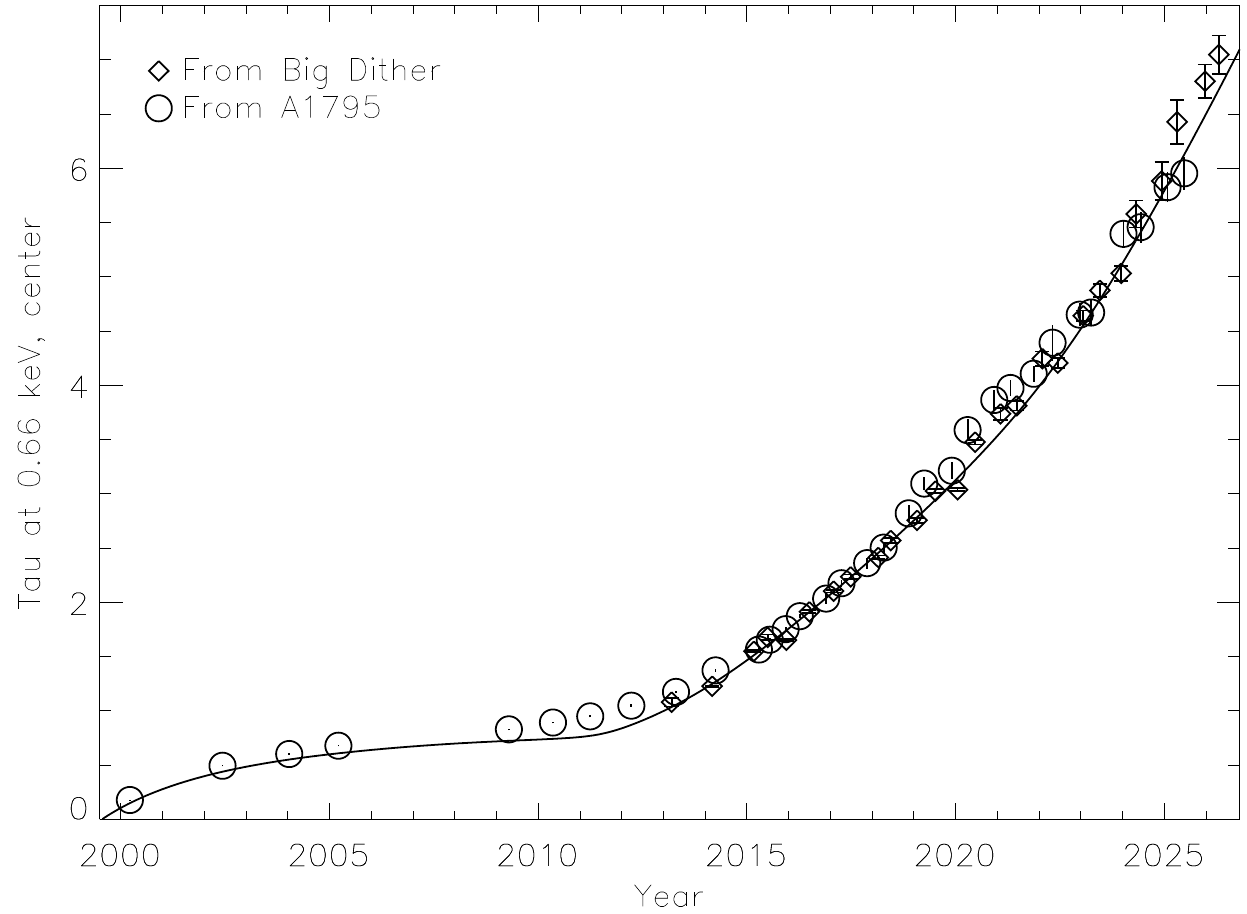}
   \includegraphics[height=6.1cm]{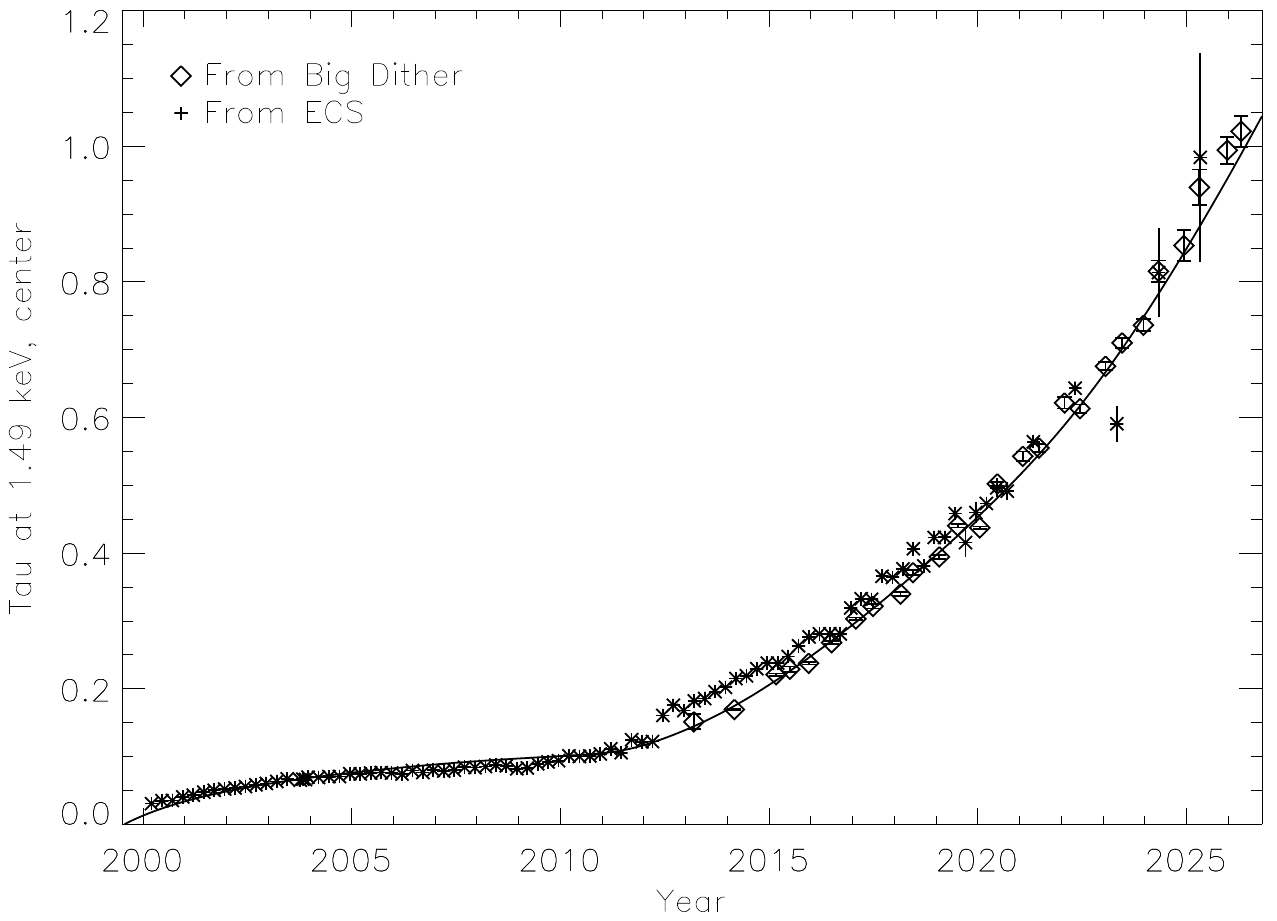}
   \end{center}
   \caption[example] 
   { \label{fig:Mrk421_optdepth} LEFT: Optical depth at 0.66~keV derived from the big dither LETGS data for the center of the ACIS-S3 compared to that derived from the A1795 data. The solid line is the prediction from the N0016 model.
   RIGHT: Optical depth at 1.49~keV derived from the big dither LETGS  data for the center of ACIS-S3 compared to that derived from the ECS data. The solid line is the prediction from the N0016 model.
}
   \end{figure}

The Mrk~421 ``big dither'' observations generally cover three broad sectors of the ACIS-S array by sweeping the dispersed LETG spectra: from rows 30-300, rows 250-520, and rows 730-1000.  Before 2014, most data came from near row 150 using a variety of blazars, with occasional observations at the center or readout rows; these were adjusted for the spatial contaminant pattern in order to estimate accumulated optical depth at the center of the detector.  The optical depths at energies of 0.66~keV and 1.49~keV are shown in Figure~\ref{fig:Mrk421_optdepth}, estimated using the optical depths of the components, extrapolated to the relevant energy using opacity models.   The optical depth at 0.66~keV agrees well with that derived from the A1795 data and the optical depth at 1.49~keV agrees well with that derived from the ECS data.

\begin{figure} [ht]
   \begin{center}
   \includegraphics[height=6.0cm]{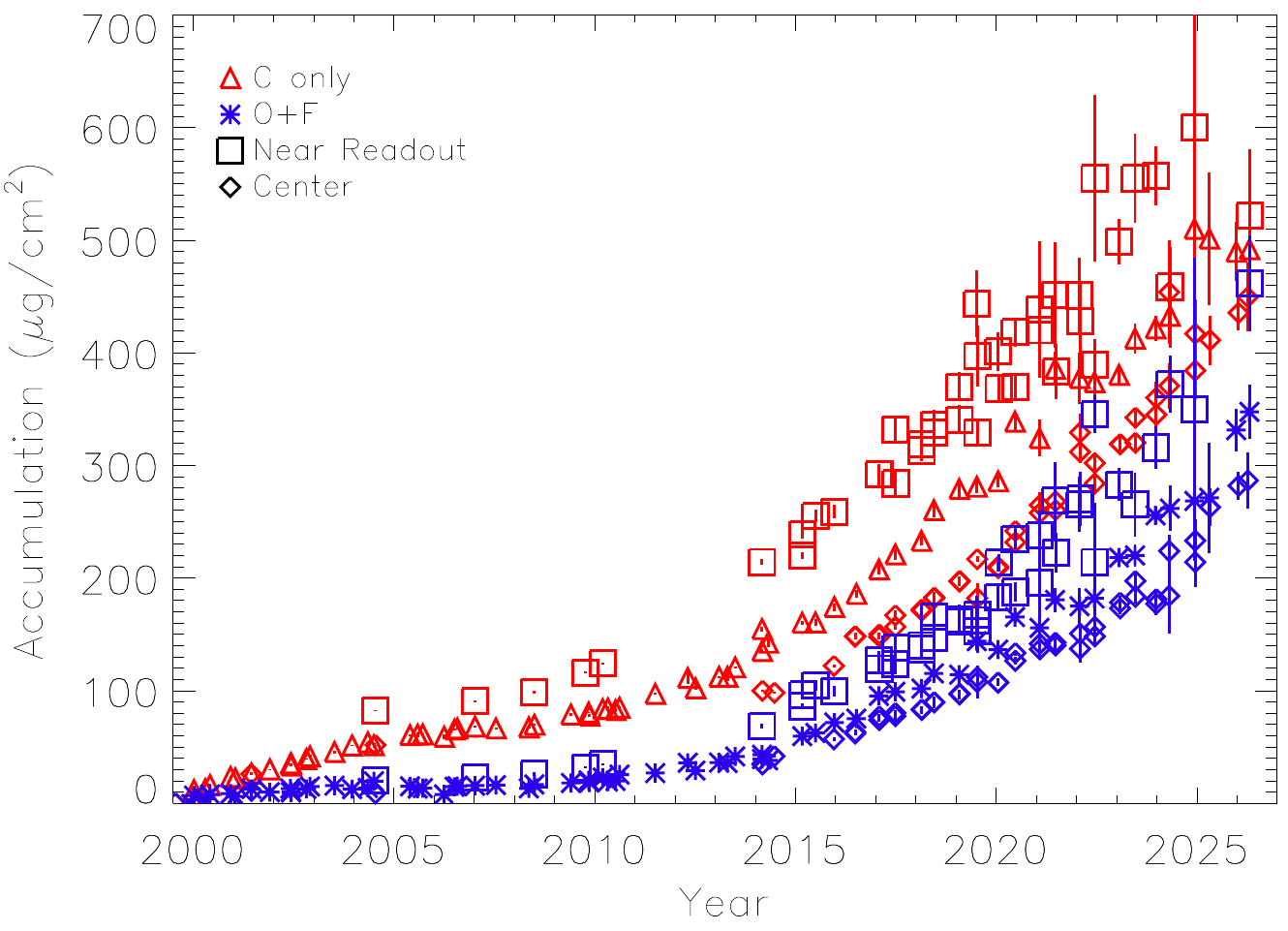}
   \includegraphics[height=6.0cm]{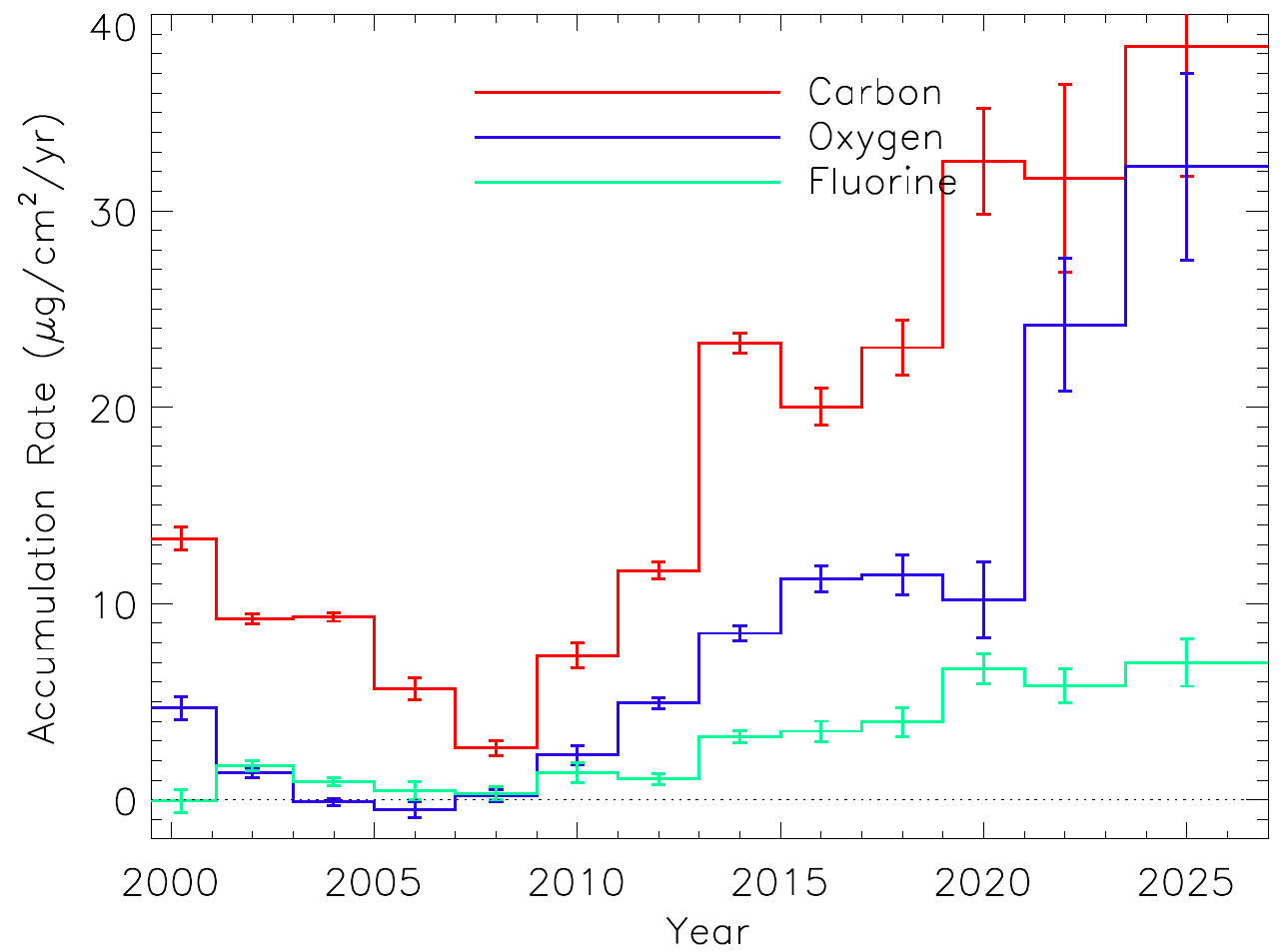}
   \end{center}
   \caption[example] 
   { \label{fig:marshall} LEFT: Areal densities of C (red) and O+F (blue) as a function of time measured from fits to Mrk~421 data.  The different symbols indicate different locations on the filter. The diamonds are in the center, the triangles are $\sim1/3$ from the edge, and the boxes are at the edge.
   RIGHT: C(red) O(blue), \& F(green) accumulation rates as a function of time in two year intervals in the middle of the S3 array. Note that the accumulation rate for O dropped to about 0 for a three year period from 2004-2006.
}
   \end{figure} 

\begin{figure} [ht]
   \begin{center}
   \includegraphics[height=10.0cm]{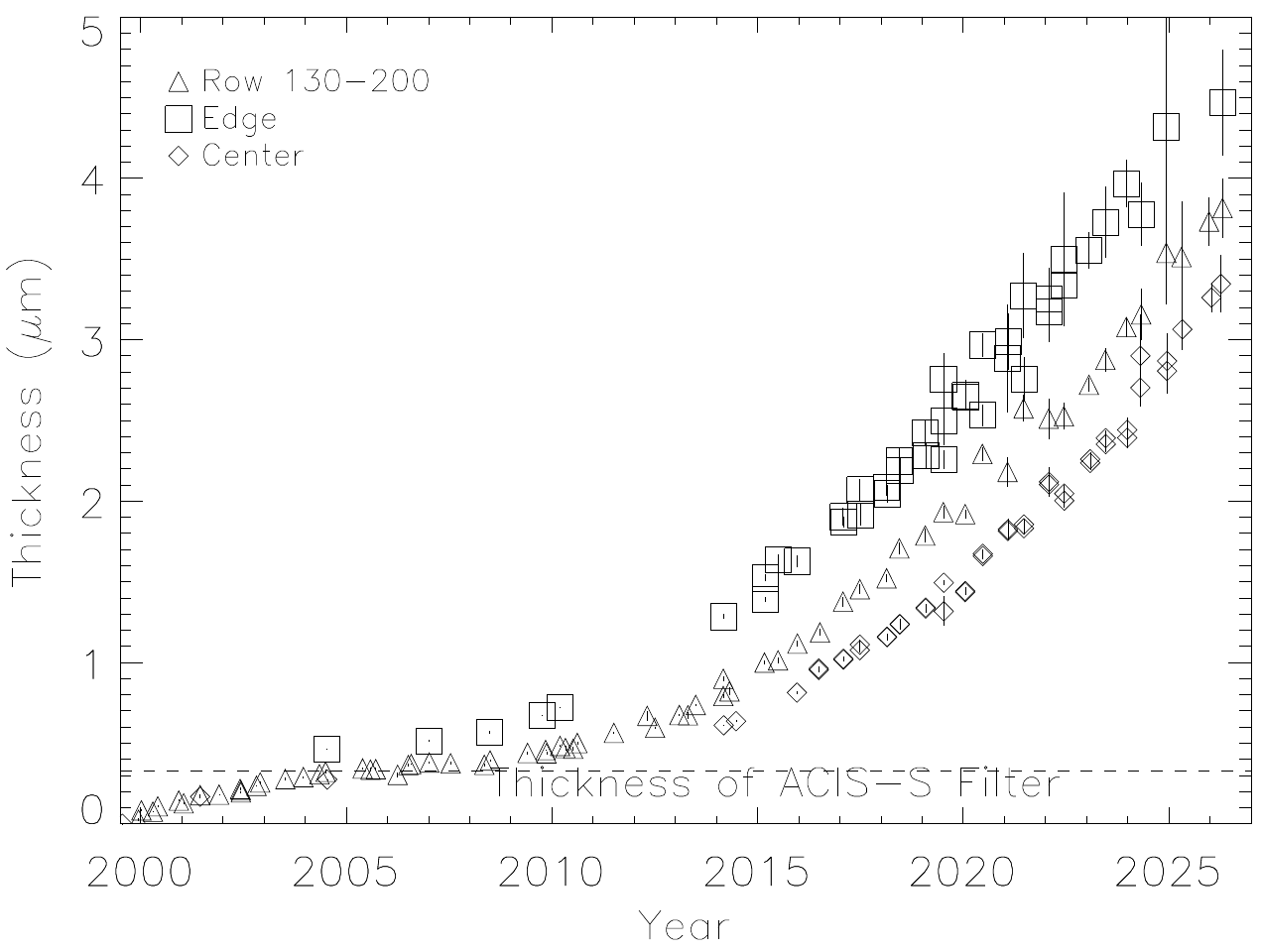}
   \end{center}
   \caption[example] 
   { \label{fig:thickness} 
   Estimated thickness of the contamination layer assuming a density of  $2.2\, {\mathrm{gm\, cm^{-3}}}$.
   The different symbols indicate different locations on the filter. The diamonds are in the center, the triangles are $\sim$1/3 from the edge, and the boxes are at the edge. All components are combined (C+O+F). 
   The dashed black line is the thickness of the ACIS-S filter. }
   \end{figure}

 The Mrk~421 data have been analyzed to measure the increase in the C, O, and F edges over the course of the mission.  The derived optical depths may be used to estimate the thickness of the contamination layer with some assumptions. For C, O, and F, the number of atoms corresponding to an optical depth of unity
 at their respective K edges
 is $9.6\times10^{17}$, $1.75\times10^{18}$, and $2.2\times10^{18}$, respectively, based on absorption cross sections, giving areal densities based on optical depths using the atomic mass of each element.  The current areal densities of C and the combination of O $+$ F shown in Figure~\ref{fig:marshall} (left) are estimated to be $\sim450$ and $\sim300$ $\mu$g cm$^{-2}$, respectively, at the center of ACIS-S, rising to $>600$ and $\sim450$ $\mu$g cm$^{-2}$ at the edges of the array.
Only total optical depths are measurable, so these were differentiated numerically to determine accumulation rates in 2 year intervals.
Figure~\ref{fig:marshall} (right) shows how the C, O, and F accumulation rates have evolved in time.  C is the dominant component in the contaminant, with O the second most important element and F the least important.  
The accumulation rate of C and O decreased from 2000 until 2008, while changes in F are barely detectable during this time interval.  After 2008, the C, O, and F accumulation rates increased significantly.
The O accumulation rate appears to be increasing linearly since 2010, while the C and F rates may have leveled off since 2020.
One important feature of Fig.~\ref{fig:marshall} (right) is that the O/C ratio dramatically increased in 2010, indicating that the composition of the contaminant changed significantly at about that time.  Meanwhile, the F/C ratio has stayed about the same since 2010.
Another significant change in the O/C ratio occurred in 2022, although the uncertainties on the measurements are relatively large.  The O and C edge measurements have become more covariant with each other as the observed counts from Mrk~421 have decreased over the mission. This can be seen in the measurements in 2020 in which the C depth increased while the O depth decreased and in 2022 in which the O depth increased and the C depth decreased (although only the increase in the O depth in 2022 is statistically significant).  We will explore alternate methods of determining these edge depths in the future.
The total areal density was converted to a thickness based on a density estimate of $2.2\, {\mathrm{gm\, cm^{-3}}}$.  This derived thickness versus time is plotted in Figure~\ref{fig:thickness}. Note that the contamination layer at the center became thicker than the ACIS OBF in $\sim$ 2006 and is now $>10\times$ thicker.

\section{CONCLUSIONS}
\label{sec:conclusions}  

The analysis of the calibration observations of E0102, A1795, \& Mrk~421 demonstrate that the N0016 contamination model is accurately modeling the increase of the contamination layer from 2022 until mid-2026 near the aimpoints on the S3 and I3 CCDs.  In particular, the E0102 \ion{Ne}{ix}~He$\alpha$~{\em r}, \ion{Ne}{x}~Ly$\alpha$, and  \ion{Mg}{ix}~He$\alpha$~{\em r} line normalizations at mid chipy on S3 and at high chipy on I3 are consistent with each other to within $\pm6\%$ from 2022-2026, except for \ion{Ne}{ix} on I3 mid chipy which has a larger scatter of $\pm9\%$.  Over the course of the entire mission, these line normalizations agree with each other within  $\pm11\%$.   The flux ratios from A1795 of the $0.5-1.0$~keV and $1.0-2.0$~keV bands at mid, low, and high chipy on the S3 CCD are consistent with each other to within $\pm3\%$ over the entire mission.  There might be a downward trend in the ratios at low and high chipy at late times but the trend is only significant at the $1.0\sigma$ level currently.  The optical depth at 0.66~keV derived from the A1795 data near the aimpoints on the S3 and I3 CCDs agrees well with the predicted optical depth from the N0016 model.  The optical depth at 0.66~keV derived from the Mrk~421 data agrees well with that predicted by the N0016 model and measured by the A1795 data and the optical depth at 1.49~keV derived from the Mrk~421 data agrees well with that predicted by the N0016 model and measured by the ECS data.  We conclude that the N0016 model is a significant improvement over previous models based on these up-to-date calibration observations.

There are some interesting trends in the data that warrant further investigation in the future as new calibration observations are obtained.  The increase in the O accumulation rate measured by the high resolution spectra of Mrk~421 starting in 2022 might indicate a change in the composition of the contaminant or a new species of contaminant, but future observations and/or new analysis techniques will be needed to confirm the significance of this result. The decrease in the E0102 \ion{Ne}{x}~Ly$\alpha$ line normalization at low chipy on I3 where the contaminant is thickest might indicate that the spatial model needs adjustment on the ACIS-I array.  Finally, the behavior of the E0102 \ion{Ne}{x}~Ly$\alpha$ line normalizations at low and high chipy on S3 are difficult to explain in terms of the spatial model of the contamination alone.  It is likely that detector effects such as CTI and gain are contributing to the scatter in these measurements and a combination of modifications to the contamination model and detector response files will be needed to fully explain these results.

\acknowledgments 
 
We thank all members of the CXC calibration and ACIS instrument teams that have supported us in this effort. P.P.P., P.W.R. and A.B. acknowledge support under NASA contract NAS8-03060 with the Chandra X-ray Center.
Support for this work was provided in part by the National Aeronautics and
Space Administration (NASA) through the Smithsonian Astrophysical Observatory (SAO)
contract SV3-73016 to MIT for support of the Chandra X-Ray Center (CXC),
which is operated by SAO for and on behalf of NASA under contract NAS8-03060.


\bibliography{spie2020} 
\bibliographystyle{spiebib} 

\end{document}